\documentclass[10pt,letterpaper]{article}
\usepackage[top=0.85in,left=2.75in,footskip=0.75in]{geometry}

\usepackage{amsmath,amssymb}

\usepackage{changepage}
\usepackage[utf8]{inputenc}
\usepackage[T1]{fontenc}

\usepackage{amsmath}
\usepackage{amssymb}
\usepackage{bm} 
\def\bm#1{{\boldsymbol{#1}}} 

\usepackage{graphicx}
\usepackage{float}     
\usepackage{placeins}  
\usepackage{pdflscape} 
\usepackage{lscape}    

\usepackage{algorithm}
\usepackage{algorithmic}

\usepackage{setspace}  
\usepackage{lineno}    
\usepackage{hyperref}  

\usepackage{natbib} 
\usepackage{textcomp,marvosym}

\usepackage{cite}

\usepackage{nameref,hyperref}

\usepackage[nopatch=eqnum]{microtype}
\DisableLigatures[f]{encoding = *, family = * }

\usepackage[table]{xcolor}

\usepackage{array}

\newcolumntype{+}{!{\vrule width 2pt}}

\newlength\savedwidth

\newcommand\thickhline{\noalign{\global\savedwidth\arrayrulewidth\global\arrayrulewidth 2pt}%
\hline
\noalign{\global\arrayrulewidth\savedwidth}}

\raggedright
\usepackage[aboveskip=1pt,labelfont=bf,labelsep=period,justification=raggedright,singlelinecheck=off]{caption}

\makeatletter
\renewcommand{\@biblabel}[1]{\quad#1.}
\makeatother

\usepackage{lastpage,fancyhdr,graphicx}
\usepackage{epstopdf}
\fancyheadoffset[L]{2.25in}
\fancyfootoffset[L]{2.25in}
\newcommand{\lorem}{{\bf LOREM}}
\newcommand{\ipsum}{{\bf IPSUM}}

\begin{document}
\vspace*{0.2in}

\begin{flushleft}
{\Large
\textbf\newline{Bayesian Joint Modeling of Zero-Inflated Longitudinal Data and Survival with a Cure Fraction: Application to AIDS Data} 
}
\newline
\\
Taban Baghfalaki\textsuperscript{1,*},
Mojtaba Ganjali\textsuperscript{2}
\\
\bigskip
\textbf{1} 
Department of Mathematics, The University of Manchester,  Manchester, UK
\\
\textbf{2} Department of Statistics, Faculty of Mathematical Sciences, Shahid Beheshti University, Tehran, Iran.
\\
\bigskip

%
%





* taban.baghfalaki@manchester.ac.uk

\end{flushleft}
\section*{Abstract}
We propose a comprehensive Bayesian joint modeling framework for zero-inflated longitudinal count data and time-to-event outcomes, explicitly incorporating a cure fraction to account for subjects who never experience the event. The longitudinal sub-model employs a flexible mixed-effects Hurdle model, with distributional options including zero-inflated Poisson and zero-inflated negative binomial, accommodating excess zeros and overdispersion common in count data. The survival component is modeled using a Cox proportional hazards model combined with a mixture cure model to distinguish cured from susceptible individuals. To link the longitudinal and survival processes, we include a linear combination of current longitudinal values as predictors in the survival model.
Inference is performed via Hamiltonian Monte Carlo, enabling efficient and robust parameter estimation. The joint model supports dynamic predictions, facilitating real-time risk assessment and personalized medicine. Model performance and estimation accuracy are validated through simulation studies. Finally, we illustrate the methodology using a real-world HIV cohort dataset, demonstrating its practical utility in predicting patient survival outcomes and supporting personalized treatment decisions. Our results highlight the benefits of integrating complex longitudinal count data with survival information in clinical research.
\\

{Keywords}: Hurdle model; Hamiltonian Monte Carlo; Longitudinal count data; Mixture model; Survival analysis.



\section{Introduction}
In many clinical and biomedical studies, traditional survival analysis assumes that all individuals will eventually experience the event of interest, such as relapse, disease progression, or death. This assumption is often unrealistic, particularly when a subset of patients is effectively cured and will never experience the event. Cure models address this limitation by explicitly incorporating a cured fraction, enabling the distinction between individuals susceptible to the event and those who are cured. These models are typically informed by long-term follow-up data \citep{farewell1982use,maller1996survival,zhao2010semiparametric} and have become essential for accurately characterizing long-term survival patterns and capturing heterogeneous treatment effects across diseases. In biomedical contexts, cured individuals are often referred to as long-term survivors. Cure models, also called long-term survival models, generally treat the study population as a mixture of two latent subgroups: those susceptible to the event and those who are not \citep{zhao2010semiparametric,zhao2012measurement,sy2000estimation,peng2000nonparametric}.\\
Longitudinal studies, which involve the repeated measurement of biomarkers or clinical parameters over time, offer rich and detailed insights into the dynamic progression of diseases. By capturing temporal changes, these studies help to elucidate how biological markers evolve in relation to disease status, treatment response, and patient outcomes. Joint modeling of longitudinal and survival data has become a widely used methodology to capture the association between dynamic biomarkers and time-to-event outcomes \citep{rizopoulos2012joint,wulfsohn1997joint}. 
 When a cure fraction exists in the population, failing to account for it in joint models can lead to biased survival estimates and inaccurate prediction.
Integrating cure fractions into joint models of longitudinal and survival data allows researchers to more precisely characterize the complex relationship between biomarker trajectories and the likelihood of cure or event occurrence. This integration is particularly valuable because it enables simultaneous modeling of both the longitudinal biomarker evolution and the time-to-event outcome, while accounting for a subset of patients who may never experience the event due to effective cure.
\\
Mixture cure models have been successfully extended within the joint modeling framework to accommodate this complexity, allowing for the identification and separate analysis of a cured subgroup alongside those susceptible to the event \citep{chen2004new,kim2013joint,yu2008individual,law2002joint,yu2004joint,pan2014joint}. Such models improve the accuracy of risk predictions and provide deeper understanding of heterogeneity in patient prognosis. Furthermore, they facilitate the study of how changes in biomarkers over time influence the probability of being cured versus experiencing the event \citep{yang2021joint}.
\\
Bayesian approaches offer distinct advantages in this setting. Their inherent flexibility supports the modeling of complex hierarchical  structures typical in longitudinal and survival studies \citep{martins2017joint,barbieri_2020,brown2003bayesian}. These features make Bayesian joint mixture cure models a powerful tool for advancing personalized medicine and improving understanding of long-term disease outcomes. To estimate these complex models, Markov Chain Monte Carlo (MCMC) algorithms are typically employed, enabling efficient approximation of posterior distributions.
\\
An additional challenge arises when longitudinal measurements exhibit zero-inflation, i.e., an excess of zeros beyond what standard count or continuous models expect. Zero-inflation commonly occurs in biomedical data due to biological absence of a marker or measurement limitations \citep{Lambert92}. Ignoring zero-inflation in joint models may lead to biased parameter estimates and poor predictive performance. To address this, \citep{baghfalaki2021approximate} proposed a Bayesian joint modeling framework that incorporates zero-inflated longitudinal data and survival outcomes, using latent variables to distinguish between structural zeros and sampling zeros while capturing the association with the survival process. This approach improves biological interpretability and enhances dynamic risk prediction.
Building on this foundation, \citep{ganjali2024joint} developed a comprehensive Bayesian framework for jointly modeling zero-inflated longitudinal measurements and survival outcomes, with a particular focus on dynamic prediction. They introduced the \textit{UHJM} R package, which implements a Gibbs sampler for hurdle joint models. The package accommodates various distributional assumptions and enables estimation of joint models for zero-inflated longitudinal data and survival times, as well as dynamic risk prediction, offering a versatile and practical tool for applied researchers.\\
This paper addresses a critical gap in joint modeling by integrating zero-inflated longitudinal data with time-to-event outcomes in the presence of a cure fraction. We employ a flexible mixed-effects hurdle model for the longitudinal submodel, allowing for zero-inflated Poisson and zero-inflated negative binomial distributions to handle excess zeros and overdispersion typical in count data. The survival component is specified through a Cox proportional hazards mixture cure model that distinguishes cured from susceptible individuals. To link the longitudinal and survival processes, we consider various functional forms including linear combinations of current values, slopes, and shared random effects. Given the model’s complexity, Bayesian inference is performed using \texttt{rstan} with Hamiltonian Monte Carlo and variational Bayes, facilitating efficient and robust estimation. This methodology extends prior research that employs MCMC-based Bayesian frameworks such as JAGS for joint modeling of zero-inflated longitudinal data \citep{ganjali2024joint} and  joint modeling of  longitudinal data and survival with cure fraction   \citep{barbieri_2020}. A key advantage of the proposed joint model is its capability to generate dynamic predictions, which facilitate real-time risk assessment and personalized clinical decision-making. 
\\
The remainder of this paper is organized as follows. Section 2 introduces the key notation and presents the proposed models, detailing the two main components of the zero-inflated joint models (ZIJMs) with a cure fraction: the zero-inflated longitudinal submodel and the time-to-event submodel incorporating the cure fraction. This section also covers the likelihood formulation and parameter estimation procedures. Section 3 focuses on dynamic prediction methodologies using ZIJMs.  Section 4 presents simulation studies designed to evaluate the performance of the proposed joint modeling approach in terms of parameter estimation, accuracy and dynamic prediction.
In Section 5, we demonstrate the application of the models to a real-world data set.
Finally, Section 6 concludes the paper with a discussion of key findings and potential directions for future research.

\section{Model Specification}
For each subject \( i \), let \( T_i^* \) denote the  true  failure time, and \( C_i \) the censoring time. The observed time-to-event is then defined as
\[
T_i = \min(T_i^*, C_i),
\]
with the event indicator \( \delta_i \) given by
\[
\delta_i = \begin{cases}
1, & \text{if } T_i^* \leq C_i \ (\text{event observed}), \\
0, & \text{if } T_i^* > C_i \ (\text{censored}).
\end{cases}
\]
Let \( Y_{ij} = Y_i(s_{ij}) \) denote the longitudinal measurement for subject \( i \) at time point \( s_{ij} \). The full observed data consist of
\[
\{T_i, \delta_i, Y_{ij} : i = 1, \ldots, N;\ j = 1, \ldots, n_i\},
\]
which represents repeated measurements and event outcomes sampled from the target population.

\subsection{Zero-inflated Longitudinal Sub-model}
We consider longitudinal count data that exhibit zero inflation. To model such data, we adopt a {hurdle model} \citep{mullahy1986specification}, which effectively accounts for the excess zeros. We favor the hurdle model due to its flexibility in handling the data-generating process. For a comparative analysis of zero-inflated models \citep{lambert1992zero} and hurdle models \citep{mullahy1986specification}, see also \citep{feng2021comparison}, where a finite mixture model is used for the response.
\\
In this study, we employ a {hurdle joint model}, where the longitudinal sub-model assumes a {zero-truncated} distribution for the positive counts. Let \( Y_i(s_{ij}) \) denote the count response for subject \( i \) at measurement time \( s_{ij} \). Its probability mass function is given by
\begin{eqnarray} \label{zi1}
P\!\left(Y_i(s_{ij}) = y_{ij} \,\middle|\, \pi^\mathcal{Z}_i(s_{ij})\right) 
&=& 
\left\{
\begin{array}{ll}
\pi^\mathcal{Z}_i(s_{ij}), & y_{ij} = 0, \\[8pt]
\left( 1 - \pi^\mathcal{Z}_i(s_{ij}) \right) 
P_{ZT}\!\left(y_{ij} \mid \bm{\vartheta}_i \right), 
& y_{ij} = 1, 2, \ldots
\end{array}
\right. \nonumber \\[8pt]
&=& 
\big[ \pi^\mathcal{Z}_i(s_{ij}) \big]^{\Delta_0(y_{ij})} 
\big[ \left( 1 - \pi^\mathcal{Z}_i(s_{ij}) \right) P_{ZT}\!\left(y_{ij} \mid \bm{\vartheta}_i \right) \big]^{1 - \Delta_0(y_{ij})}.
\end{eqnarray}
Here, \( P_{ZT}\!\left( \cdot \mid \bm{\vartheta}_i \right) \) denotes the zero-truncated probability mass function with parameter vector \( \bm{\vartheta}_i \). The function \( \Delta_0(\cdot) \) is the Dirac indicator function for zero:
\[
\Delta_0(y) = 
\begin{cases}
1, & y = 0, \\
0, & y \neq 0,
\end{cases}
\]
such that \( \pi^\mathcal{Z}_i(s) \in [0,1] \) represents the probability of a {structural zero}, and positive counts occur with probability \( 1 - \pi^\mathcal{Z}_i(s) \), following the zero-truncated distribution.
For this structure, a wide range of distributions from the exponential dispersion family, as well as semi-continuous distributions, can be considered \citep{ganjali20252}, and our methodology can be adapted accordingly based on the framework presented therein.
In this paper, we focus on the zero-inflated Poisson (ZIP) and zero-inflated negative binomial (ZINB) models, a choice supported by empirical evidence demonstrating their strong performance in real-world applications \citep{baghfalaki2021approximate, ganjali2024joint}.\\
For the zero-truncated negative binomial (ZTNB) distribution, we consider
$$P_{ZTNB}\left(y_{ij} \mid \kappa_i(s_{ij}), r \right) =\displaystyle\frac{
\frac{\Gamma(y_{ij} + r)}{y_{ij}! \, \Gamma(r)} \, \kappa_i(s_{ij})^r \, (1 - \kappa_i(s_{ij}))^{y_{ij}}
}{
1 - \kappa_i(s_{ij})^r
},~y_{ij}=1,2,\cdots,$$
where
\( \kappa_i(s) \in (0,1) \) governs the success probability, and \( r > 0 \) is a dispersion parameter controlling the overdispersion.
The ZTNB distribution is parameterized in terms of \( \kappa_i(s) \), which is related to the mean \( \lambda_i(s) \) through the transformation:
\[
\lambda_i(s) = \frac{r \left(1 - \kappa_i(s)\right)}{\kappa_i(s)}.
\]
Using this reparameterization, the mean and variance of a negative binomial distribution with parameters \(\kappa_i(s)\) and \(r\) are given by \(\lambda_i(s)\) and 
$\lambda_i(s) \left(1 + \frac{\lambda_i(s)}{r} \right)$,
respectively,
yielding a dispersion index of \( 1 + \lambda_i(s)/r \), which exceeds 1 in the presence of over-dispersion. We denote this model by \( Y_i(s) \sim \text{ZINB}(\pi^\mathcal{Z}_i(s), \lambda_i(s), r) \), where \( \pi^\mathcal{Z}_i(s) \) is the probability of a structural zero and \( \lambda_i(s) \) is the mean of the underlying negative binomial component.
\\
For the zero-truncated Poisson (ZTP) distribution parameterized by the mean \(\lambda_i(s_{ij})\), the probability mass function (PMF) for \(y_{ij} = 1, 2, \ldots\) is given by
$$P_{ZTP}\left(y_{ij} \mid \lambda_i(s_{ij}) \right) = \frac{
e^{-\lambda_i(s_{ij})} \frac{\lambda_i(s_{ij})^{y_{ij}}}{y_{ij}!}
}{
1 - e^{-\lambda_i(s_{ij})}
}, \quad y_{ij} = 1, 2, \cdots.$$
To characterize subject-specific variation in both the mean count and the probability of excess zeros, we specify a mixed-effects regression framework as follows.
Let \( \bm{x}_{1i}(s) \in \mathbb{R}^{p_1} \) and \( \bm{x}_{2i}(s) \in \mathbb{R}^{p_2} \) denote covariate vectors used to model the mean \( \lambda_i(s) \) and the structural zero probability \( \pi^\mathcal{Z}_i(s) \), respectively. These covariate sets may overlap. The longitudinal sub-model is specified as a Hurdle-type generalized linear mixed model:
\begin{equation} \label{zi2}
\begin{aligned}
\eta_i^\lambda(s) &= \log \lambda_i(s) = \bm{x}_{1i}(s)^\top \bm{\beta}_1 + \bm{z}_{1i}(s)^\top \bm{u}_i, \\
\eta_i^\mathcal{Z}(s) &= \text{logit}(\pi^\mathcal{Z}_i(s)) = \bm{x}_{2i}(s)^\top \bm{\beta}_2 + \bm{z}_{2i}(s)^\top \bm{b}_i,
\end{aligned}
\end{equation}
where \( \bm{\beta}_1 \in \mathbb{R}^{p_1} \) and \( \bm{\beta}_2 \in \mathbb{R}^{p_2} \) are vectors of fixed effects, while \( \bm{u}_i \in \mathbb{R}^{q_1} \) and \( \bm{b}_i \in \mathbb{R}^{q_2} \) are subject-specific random effects associated with the design vectors \( \bm{z}_{1i}(s) \) and \( \bm{z}_{2i}(s) \), respectively.
\\
The random effects are assumed to follow multivariate normal distributions, specifically \( \bm{u}_i \sim \mathcal{N}(\bm{0}, \boldsymbol{D}_u) \) and \( \bm{b}_i \sim \mathcal{N}(\bm{0}, \boldsymbol{D}_b) \).\\
The model in \eqref{zi2}, often referred to as a two-part regression model, is widely used in the analysis of longitudinal zero-inflated count data \citep{alfo2010two, maruotti2011two, farewell2017two,baghfalaki2021approximate, ganjali2024joint}.

\subsection{Survival Sub-model with Cure Fraction}\label{mcm}
For the survival sub-model incorporating a cure fraction, we consider   the classical mixture cure model \citep{boag1949maximum}, which assumes that the population comprises a mixture of susceptible and non-susceptible (cured) individuals. \\
Let $\mathcal{C}$ be the indicator of uncured status, where $\mathcal{C}=1$ represents an uncured subject and $\mathcal{C}=0$ represents a cured subject. The mixture cure model assumes that the survival function for the whole population is defined as a mixture of survival functions defined by:
\begin{equation}\label{eq:mcm}
S\left( t \right) = \pi^\mathcal{C}+  \left( 1- \pi^\mathcal{C} \right)   S\left( t \vert \mathcal{C}=1 \right)
\end{equation}
where $S\left( t \vert \mathcal{C}=1 \right)$ is the proper survival function associated with a proportion $\pi^\mathcal{C}$ of uncured subjects, while the one associated with the proportion $(1-\pi^\mathcal{C})$ of cured subjects is implicitly fixed at 1 whatever $t\in\mathbb{R}^{+}$. The proportion $\pi^\mathcal{C}$ of the mixture is usually not considered as a parameter as in classical mixture models but is rather modeled based on the available data. The probability $\pi^\mathcal{C}_i$ of being uncured for the \textit{i}-th subject is classically modeled through a logistic regression model, called the "incidence model", such as:
\begin{eqnarray}\label{eq:mcm_incidence}
\text{logit}(\pi^\mathcal{C}_i) = \text{logit}(P\left(\mathcal{C}_i=1\vert \boldsymbol{W_{1i}}=\boldsymbol{w_{1i}}\right)) =\boldsymbol{w_{1i}}^{\top}\boldsymbol{\xi}_1,
\end{eqnarray}
with $\boldsymbol{\xi}_1$ being an $r_1$-dimensional parameter vector associated with the time-independent covariates $\boldsymbol{w_{1i}}$, where the first component of both vectors corresponds to the intercept of the logistic regression model.
The survival sub-model is only defined for uncured fraction and also depends on the shared association:  
\begin{eqnarray}\label{eq:MCM_haz}
h\left(t\vert \mathcal{C}_i=1 \right) & = & h_0(t)\exp\left( \boldsymbol{w_{2}}^{\top}\boldsymbol{\xi}_2 + \alpha_1 \eta^\mu(t) + \alpha_2 \eta^\pi(t) \right),
\end{eqnarray}
where $\boldsymbol{w_{2i}}$ is an $r_2$-dimensional vector of time-independent covariates associated with the parameter vector $\boldsymbol{\xi}_2$, and $h_0(t)$ denotes the baseline hazard function.
The two vectors of baseline covariates $\boldsymbol{w_{1i}}$ and $\boldsymbol{w_{2i}}$ may or may not overlap with each other. 
To capture the association between the zero-inflated (ZI) longitudinal outcome and the survival model, we include the term $\alpha_1 \eta^\mu(t) + \alpha_2 \eta^\pi(t)$, which is a linear combination of the log-mean predictor $\eta^\mu(t)$ and the logit-zero predictor $\eta^\pi(t)$. Here, $\alpha_1$ and $\alpha_2$ are the association parameters. For alternative specifications of the association structure, see \citep{ganjali2024joint}.\\
The associated survival function is then defined by:
\begin{eqnarray*}\label{eq_surv_JLSCM}
S\left( t\vert \mathcal{C}_i=1 \right) & = & \exp\left\{ -\int_{0}^{t}  h\left(s\vert \mathcal{C}_i=1 \right) ds \right\}.
\end{eqnarray*}
Hereafter, let $\boldsymbol{C}=\{\mathcal{C}_1,\cdots,\mathcal{C}_n\}$.\\
As our methodology is not restricted to a specific form of the baseline hazard, we also consider apiecewise-constant baseline hazard function defined as
\[
h_{0}(t) = \sum_{q=1}^Q h^*_q \, I(\nu_{q-1} < t \leq \nu_q),
\]
where $0 = \nu_0 < \nu_1 < \cdots < \nu_Q$ represents a partition of the time scale, with $\nu_Q$ exceeding the largest observed time, and $h^*_q$ denotes the constant hazard within the interval $(\nu_{q-1}, \nu_q)$.\\
We now consider two general joint models with a cure fraction: the joint model of zero-inflated negative binomial longitudinal outcomes and survival with a cure fraction, and its zero-inflated Poisson counterpart. We denote these joint models as $\text{ZINBJMCF} (\pi^\mathcal{Z}_i(s), \lambda_i(s), r)$ and $\text{ZIPJMCF} (\pi^\mathcal{Z}_i(s), \lambda_i(s))$, respectively, where $\pi^\mathcal{Z}_i(s)$ and $\lambda_i(s)$ are defined in \eqref{zi2}.  
Since the survival sub-model is identical in both joint models, we omit its notation in our expressions. Nevertheless, in both models, equations \eqref{eq:mcm_incidence} and \textbf{\eqref{eq:MCM_haz}} describe the survival sub-model with a cure fraction.

\subsection{Bayesian inference}\label{pd}
The posterior distribution is given by $$\pi\left(\boldsymbol{\theta},\boldsymbol{C},\boldsymbol{u},\boldsymbol{b} \vert \boldsymbol{y},\boldsymbol{t},\boldsymbol{\delta}\right) \propto \mathcal{L}\left( \boldsymbol{y},\boldsymbol{t},\boldsymbol{\delta} \vert \boldsymbol{C}, \boldsymbol{u},\boldsymbol{b},\boldsymbol{\theta} \right) \pi\left(\boldsymbol{C},\boldsymbol{u},\boldsymbol{b},\boldsymbol{\theta}\right)$$ where $\mathcal{L}$ denotes the likelihood and 
$\boldsymbol{\theta}=\{ \boldsymbol{\beta}_1, \boldsymbol{\beta}_2, \boldsymbol{D}_u,\boldsymbol{D}_b,\alpha_1,\alpha_2, \boldsymbol{\xi}_1, \boldsymbol{\xi}_2, r  \}$ 
is the parameter vector.
 Assuming that $\pi\left(\boldsymbol{C},\boldsymbol{u},\boldsymbol{b},\boldsymbol{\theta}\right)=\pi\left(\boldsymbol{u},\boldsymbol{b}\vert\boldsymbol{C},\boldsymbol{\theta}\right)\pi\left(\boldsymbol{C}\vert\boldsymbol{\theta}\right)\pi\left(\boldsymbol{\theta}\right)$, the posterior distribution is: 
 \begin{equation}\label{jointpost}
\begin{aligned}
\pi(\boldsymbol{\theta}, \boldsymbol{C}, \boldsymbol{b}, \boldsymbol{u} \mid \boldsymbol{y}, \boldsymbol{t}, \boldsymbol{\delta}) 
&\propto 
\prod_{i=1}^{n} 
\Bigg[
\underbrace{
\prod_{j=1}^{n_i} 
P\!\left(Y_i(s_{ij}) = y_{ij} \,\middle|\, \pi^\mathcal{Z}_i(s_{ij})\right)
}_{\text{longitudinal}} \\
&\quad \times 
\underbrace{
\phi(\boldsymbol{u}_i \mid \boldsymbol{0}, \boldsymbol{D}_u)
\phi(\boldsymbol{b}_i \mid \boldsymbol{0}, \boldsymbol{D}_b)
}_{\text{random effects}} \\
&\quad \times 
\underbrace{
\bigl[\pi^{\mathcal{C}}_i h(t_i \mid \mathcal{C}_i) S(t_i \mid \mathcal{C}_i)\bigr]^{\delta_i \mathcal{C}_i} 
}_{\text{observed events in cured}} \\
&\quad \times 
\underbrace{
\bigl[\pi^{\mathcal{C}}_i S(t_i \mid \mathcal{C}_i)\bigr]^{(1-\delta_i)\mathcal{C}_i} 
}_{\text{censored in cured}} \\
&\quad \times 
\underbrace{
\bigl[1 - \pi^{\mathcal{C}}_i\bigr]^{(1-\delta_i)(1-\mathcal{C}_i)}
}_{\text{uncured}} 
\Bigg]
\pi(\boldsymbol{\theta}).
\end{aligned}
\end{equation}
where $\phi(\cdot)$ denotes the normal density of the subject-specific random effects, and $P\!\left(Y_i(s_{ij}) = y_{ij} \,\middle|\, \pi^\mathcal{Z}_i(s_{ij})\right)$ is given in equation~\eqref{zi1}. \\
For the components of $\bm{\theta}$, we assign independent weakly informative (or non-informative) prior distributions to all unknown model parameters to reflect minimal prior knowledge and allow the data to primarily drive the inference. Specifically, the priors are defined as follows:
\[
\begin{aligned}
&\bm{\beta}_k  \sim \mathcal{N}_{p_k}(\bm{\mu}_\beta, \bm{\Sigma}_\beta), \quad k=1,2,\\
&\bm{\xi}_k  \sim \mathcal{N}_{r_k}(\bm{\mu}_\xi, \bm{\Sigma}_\xi), \quad k=1,2,\\
&\alpha_k \sim \mathcal{N}(\mu_\alpha, \sigma^2_\alpha), \quad k=1,2,\\
& h^*_q\sim \mathrm{Gamma}(a_q,b_q), \quad q=1,\ldots,Q,\\
& r \sim \mathrm{Gamma}(a_r,b_r) \quad \text{(dispersion parameter of the ZINB).}
\end{aligned}
\]
For the covariance of the random effects, we consider
\[
\boldsymbol{D}_k = \boldsymbol{\Sigma}_k \mathbf{R}_k \boldsymbol{\Sigma}_k,
\]
where $\boldsymbol{\Sigma}_k = \mathrm{diag}(\sigma_{k,1}, \ldots, \sigma_{k,q_k})$ is a diagonal matrix of standard deviations, and $\mathbf{R}_k$ is the $q_k \times q_k$ correlation matrix with the prior
\[
\mathbf{R}_k \sim \mathrm{LKJ}(\varsigma_k), \quad k=1,2.
\]
The LKJ prior \citep{lewandowski2009generating} is a flexible prior for correlation matrices commonly used in Bayesian modeling. The shape parameter $\varsigma_k$ controls the concentration of the prior:  
\(\varsigma_k = 1\) corresponds to a uniform prior over all valid correlation matrices,  
\(\varsigma_k > 1\) concentrates mass near the identity matrix (i.e., weaker correlations), and  
\(\varsigma_k < 1\) favors stronger correlations.  
In this model, we set \(\varsigma_k = 2\) to favor mild correlations among the random effects while maintaining flexibility.  
\\
For the case of one-dimensional random effects, we use a weakly informative Cauchy prior for the variance of the random effect.

\subsection{Posterior computation}
Computing expectations under the joint posterior \eqref{jointpost} is analytically intractable, necessitating the use of approximation methods. Traditional MCMC algorithms \citep{metropolis1953equation,hastings1970monte} are often inefficient for this joint model due to slow convergence and poor scaling in high-dimensional parameter spaces. To address these challenges, we employ Hamiltonian Monte Carlo (HMC) and {Variational Bayes (VB)} using the \texttt{rstan} package \citep{guo2020package}, which provide efficient and scalable inference for complex Bayesian models. All code for the simulation study, as well as the application, is available at \url{https://github.com/tbaghfalaki/HJMCF}.
\\
HMC addresses these inefficiencies by introducing momentum variables and simulating Hamiltonian dynamics to propose distant, energy-conserving moves through the parameter space. This improves both exploration and acceptance rates. As demonstrated in \citep{neal2011mcmc} and further elaborated by \citep{betancourt2017conceptual}, HMC mitigates the random walk behavior of classical MCMC and yields more efficient sampling, particularly for models with complex geometry. With accessible implementations such as {Stan} \citep{carpenter2017stan}, HMC has become a standard tool for scalable and reliable Bayesian computation.
\\
In our joint model, HMC is employed to efficiently sample the continuous parameters, namely $\boldsymbol{\theta}, \boldsymbol{u}, \boldsymbol{b}$. Since HMC requires continuously differentiable densities, the discrete mixture components $\boldsymbol{C}$ are handled separately via  {Gibbs updates}, conditional on the current continuous parameters. This hybrid scheme alternates between:
\begin{enumerate}
    \item Sampling $\boldsymbol{C} \sim \pi(\boldsymbol{C} \mid \boldsymbol{u}, \boldsymbol{b}, \boldsymbol{\theta}, \boldsymbol{y}, \boldsymbol{t}, \boldsymbol{\delta})$ using categorical conditional probabilities.
    \item Sampling $(\boldsymbol{\theta}, \boldsymbol{u}, \boldsymbol{b})$ using HMC conditional on the current $\boldsymbol{C}$.
\end{enumerate}
This approach preserves the exactness of HMC for continuous parameters while respecting the discrete nature of the latent classes, implemented via HMC using the {No-U-Turn Sampler (NUTS)} \citep{hoffman2014no}.
\\
Variational Bayes \citep{Blei2017} provides a computationally efficient alternative to MCMC for approximating the posterior distribution \eqref{jointpost}. In VB, the joint posterior $\pi\left(\boldsymbol{\theta}, \boldsymbol{C}, \boldsymbol{b}, \boldsymbol{u} \,\big\vert\, 
\boldsymbol{y}, \boldsymbol{t}, \boldsymbol{\delta} \right)$ is approximated by a factorized distribution
\[
q(\boldsymbol{\theta}, \boldsymbol{u}, \boldsymbol{b}, \boldsymbol{C}) = 
q(\boldsymbol{\theta}) \, q(\boldsymbol{u}, \boldsymbol{b}) \, q(\boldsymbol{C}),
\]
where the approximation targets the posterior conditioned on the observed data \(\boldsymbol{y}, \boldsymbol{t}, \boldsymbol{\delta}\).  
This is achieved by minimizing the Kullback–Leibler (KL) divergence
\begin{align}
\text{KL}\Big(&q(\boldsymbol{\theta}, \boldsymbol{u}, \boldsymbol{b}, \boldsymbol{C}) 
\,\big\|\, 
\pi(\boldsymbol{\theta}, \boldsymbol{C}, \boldsymbol{b}, \boldsymbol{u} \mid 
\boldsymbol{y}, \boldsymbol{t}, \boldsymbol{\delta})\Big) \notag \\
&= \int q(\boldsymbol{\theta}, \boldsymbol{u}, \boldsymbol{b}, \boldsymbol{C}) \, 
\log \frac{
q(\boldsymbol{\theta}, \boldsymbol{u}, \boldsymbol{b}, \boldsymbol{C})
}{
\pi(\boldsymbol{\theta}, \boldsymbol{C}, \boldsymbol{b}, \boldsymbol{u} \mid 
\boldsymbol{y}, \boldsymbol{t}, \boldsymbol{\delta})
} \, 
d\boldsymbol{\theta} \, d\boldsymbol{u} \, d\boldsymbol{b} \, d\boldsymbol{C}.
\end{align}
where continuous parameters are typically modeled with Gaussian distributions and discrete components with categorical distributions. Each factor is chosen from a convenient parametric family to facilitate computation.
\\
The goal of VB is to find the factorized distribution that best approximates the posterior by maximizing the Evidence Lower Bound (ELBO), which balances fidelity to the likelihood with the complexity of the variational distribution. Expectations in the ELBO, often intractable for complex models, are typically approximated using stochastic gradient methods or Monte Carlo integration \citep{Kingma2014,Ranganath2014}.  
\\
Under the mean-field assumption, variational factors are updated iteratively using coordinate ascent: each factor is optimized while holding others fixed, ensuring monotonic increase of the ELBO and convergence to a local optimum \citep{Jordan1999}. This makes VB particularly suitable for initial exploration, large datasets, or hyperparameter tuning, while HMC remains the preferred method for final inference in complex hierarchical models \citep{Ormerod2010,Blei2017}.

\section{Dynamic prediction in ZIJM with cure Fraction}
Dynamic prediction (DP) has garnered significant attention in recent years, particularly through the joint modeling of longitudinal measurements and time-to-event outcomes \citep{rizopoulos2011dynamic}.
Dynamic prediction (DP) refers to estimating the probability that an event will occur within a future time window $(s, s + t]$, conditional on the longitudinal measurements up to time $s$, denoted by 
\[
\bm{\mathcal{Y}}_i(s) = \{Y_i(s_{ij}) : 0 \le s_{ij} \le s,\ j = 1, \ldots, n_i\},
\]
for subject $i$ who is event-free at time $s$. Let the training dataset consist of a random sample 
\[
\bm{\mathcal{D}} = \{\bm{Y}_i, T_i, \delta_i;\ i = 1, \ldots, N\},
\]
where $T_i$ is the observed event or censoring time, and $\delta_i$ is the event indicator. In clinical applications and for model validation, it is particularly important to predict risk for new subjects who were not part of the training sample.
For each subject $i$ in the new dataset, referred to as the **validation set**, we have access to their longitudinal measurements $\bm{\mathcal{Y}}_i(s)$ and a vector of available explanatory variables $\bm{\mathcal{X}}_i(s)$, defined as
\[
\bm{\mathcal{X}}_i(s) = \{\bm{x}_{1i}(s), \bm{x}_{2i}(s), \bm{w}_{1i}, \bm{w}_{2i} : 0 \le s_{ij} \le s,\ j = 1, \ldots, n_i\}.
\]
The individual dynamic risk prediction at time $s$ for a prediction window of length $t$ is then given by
\begin{eqnarray}\label{dp}
\pi_i(s + t \mid s)  & = &\mathrm{P}\left(s \leq T_i^*<s+t \mid T_i^*>s, \bm{\mathcal{Y}}_i(s), \boldsymbol{\mathcal { X }}_i(s) ; \bm{\theta} \right) \\\nonumber
 & = & 1 -  \mathrm{P}\left(T_i^* > s + t \mid T_i^* > s,\ \bm{\mathcal{Y}}_i(s),\ \bm{\mathcal{X}}_i(s) ; \bm{\theta}\right),
\end{eqnarray}
where $\boldsymbol{\theta}$ denotes the vector of all unknown model parameters, and $T_i^*$ is the true (possibly unobserved) event time for subject $i$ \citep{ganjali2024joint}. To estimate this quantity under the cure model framework, we consider the two previously discussed joint models, which can be described as follows.
\subsection{Dynamic prediction Formulation}
Let $\bm{v}_i=(\boldsymbol{u}_i, \boldsymbol{b}_i)$ denote the vector of subject-specific random effects. The conditional survival probability can then be expressed as:
\begin{align}\label{dpf_cure}
1-\pi_i(s+t \mid s) &= \mathrm{P}\Big(T_i^* > s + t \,\Big|\, 
T_i^* > s, \bm{\mathcal{Y}}_i(s), \bm{\mathcal{X}}_i(s); \bm{\theta} \Big) \notag\\
&= \int 
P(T_i^* > s + t \mid T_i^* > s, \bm{\mathcal{Y}}_i(s), \bm{\mathcal{X}}_i(s), \bm{v}_i; \bm{\theta}) \notag\\
&\quad \times 
P(\bm{v}_i \mid T_i^* > s, \bm{\mathcal{Y}}_i(s), \bm{\mathcal{X}}_i(s); \bm{\theta}) \, d\bm{v}_i \notag\\
&= \int 
\frac{S(s + t \mid \bm{\mathcal{Y}}_i(s), \bm{\mathcal{X}}_i(s), \bm{v}_i; \bm{\theta})}
     {S(s \mid \bm{\mathcal{Y}}_i(s), \bm{\mathcal{X}}_i(s), \bm{v}_i; \bm{\theta})} \notag\\
&\quad \times 
P(\bm{v}_i \mid T_i^* > s, \bm{\mathcal{Y}}_i(s), \bm{\mathcal{X}}_i(s); \bm{\theta}) \, d\bm{v}_i \notag\\
&= \int 
\frac{ \pi_i^{\mathcal{C}} +(1 - \pi_i^{\mathcal{C}}) 
      S(s + t \mid \mathcal{C}_i=1, \bm{v}_i, \bm{\mathcal{Y}}_i(s), \bm{\mathcal{X}}_i(s); \bm{\theta})}
     { \pi_i^{\mathcal{C}} +(1 - \pi_i^{\mathcal{C}})
      S(s \mid  \mathcal{C}_i=1,\bm{v}_i, \bm{\mathcal{Y}}_i(s), \bm{\mathcal{X}}_i(s); \bm{\theta})} \notag\\
&\quad \times 
P(\bm{v}_i \mid T_i^* > s, \bm{\mathcal{Y}}_i(s), \bm{\mathcal{X}}_i(s); \bm{\theta}) \, d\bm{v}_i.
\end{align}
To compute this probability, the parameter vector \( \bm{\theta} \) must first be inferred from the training dataset \( \bm{\mathcal{D}} \), resulting in an estimate \( \hat{\bm{\theta}} \). The dynamic prediction in equation~\eqref{dpf_cure} can then be evaluated using either a first-order Laplace approximation or a Monte Carlo integration technique, as described in \citep{rizopoulos2011dynamic}.
The first-order approximation of ${\mathbb{\pi}} _{i}(s+t|s)$ is as follows:
\begin{eqnarray}\label{ap20}
\hat{\mathbb{\pi}} _{i}(s+t|s)  =1-
\frac{ \hat{\pi}_i^{\mathcal{C}}+(1 - \hat{\pi}_i^{\mathcal{C}})  S(s + t \mid \mathcal{C}_i=1, \hat{\bm{v}}_i,\bm{\mathcal{Y}}_i(s),\ \bm{\mathcal{X}}_i(s); \hat{\bm{\theta}})}{ \hat{\pi}_i^{\mathcal{C}}+(1 - \hat{\pi}_i^{\mathcal{C}})  S(s \mid \mathcal{C}_i=1, \hat{\bm{v}}_i,\bm{\mathcal{Y}}_i(s),\ \bm{\mathcal{X}}_i(s); \hat{\bm{\theta}})}
\end{eqnarray}
where \( \hat{\bm{v}}_i \) denotes the empirical mean of the conditional distribution 
\(
(\bm{v}_i \mid T_i^* > s, \bm{\mathcal{Y}}_i(s), \bm{\mathcal{X}}_i(s); \hat{\bm{\theta}}),
\)
and \( \hat{\pi}_i^{\mathcal{C}} \) represents the estimated incidence probability as defined in equation~\eqref{eq:mcm_incidence}.\\
For the Monte Carlo approximation in the cure model, one proceeds as follows. Draw \( \mathcal{J} \) samples \( \bm{\theta}^{(j)} \), \( j=1,\ldots,\mathcal{J} \), from the posterior distribution of parameters based on the training data. For each \( \bm{\theta}^{(j)} \), sample \( \bm{v}_i^{(j)} \) from the posterior distribution of random effects:
\[
(\bm{v}_i \mid T_i^* > s, \bm{\mathcal{Y}}_i(s), \bm{\mathcal{X}}_i(s); \bm{\theta}^{(j)}).
\]
Then compute (for $j=1,\ldots,\mathcal{J}$)
\begin{equation}\label{dp_mc_cure}
\hat{\pi}_i^{(j)}(s+t \mid s) = 1-
\frac{  \hat{\pi}_i^{\mathcal{C}(j)}+(1 - \hat{\pi}_i^{\mathcal{C}(j)}) S(s+t \mid \mathcal{C}_i=1, \bm{v}_i^{(j)}, \bm{\mathcal{Y}}_i(s), \bm{\mathcal{X}}_i(s); \bm{\theta}^{(j)}) }
     {  \hat{\pi}_i^{\mathcal{C}(j)}+(1 - \hat{\pi}_i^{\mathcal{C}(j)}) S(s \mid \mathcal{C}_i=1, \bm{v}_i^{(j)}, \bm{\mathcal{Y}}_i(s), \bm{\mathcal{X}}_i(s); \bm{\theta}^{(j)}) },
\end{equation}
where \( \hat{\pi}_i^{\mathcal{C}(j)} \) is the cure incidence probability computed at \( \bm{\theta}^{(j)} \).\\
The collection \( \{\hat{\pi}_i^{(j)}(s+t \mid s) : j=1,\ldots,\mathcal{J}\} \) forms a posterior sample for the dynamic prediction at landmark time \( s \) and prediction window \( t \) for individual \( i \). Summary measures such as the posterior mean, credible intervals, and standard errors can be derived for inference.

\subsection{Measures of Prediction Accuracy}
Assessing the predictive performance of a model is a crucial step in the development and validation of prognostic tools \citep{steyerberg2009applications}. Two commonly used metrics for this purpose are the time-dependent Area Under the Receiver Operating Characteristic Curve (AUC) and the time-dependent Brier Score (BS) \citep{blanche2015quantifying}. These metrics evaluate a model’s ability to discriminate and calibrate individual risk predictions over time.\\
The time-dependent AUC quantifies the model’s discriminative ability—its capacity to distinguish between individuals who will experience the event within a specified prediction window and those who will not. Let \( \pi_i(s, t) \) denote the predicted risk for subject \( i \) at landmark time \( s \) over the prediction horizon \( t \). The time-dependent AUC is defined as
\[
\text{AUC}(s, t) = P\left( \pi_i(s, t) > \pi_j(s, t) \,\middle|\, D_i(s, t) = 1,\, D_j(s, t) = 0,\, T_i > s,\, T_j > s \right),
\]
where \( D_i(s, t) = \mathbb{I}\{ s < T_i \le s + t \} \) indicates whether subject \( i \) experiences the event within the interval \( (s, s + t] \).\\
In contrast, the BS measures the overall accuracy of predicted probabilities by calculating the mean squared difference between predicted risks and observed outcomes. It reflects both the calibration and sharpness of the predictions and is given by
\[
BS(s, t) = \mathbb{E} \left[ \left( D_i(s, t) - \pi_i(s, t) \right)^2 \,\middle|\, T_i > s \right].
\]
Both AUC and BS depend on the choice of the landmark time \( s \) and prediction window \( t \), which may limit interpretability when multiple time points are of interest.\\
To summarize predictive performance over a range of landmark times, integrated measures such as the integrated AUC (iAUC) and integrated Brier Score (iBS) have been proposed. These integrated 
 measures summarize the time-dependent AUC over a landmark time interval \([0, t_{\max}]\) by integrating it with appropriate weighting functions that reflect the distribution of event times. Formally, the integrated AUC is defined as
\[
iAUC(t_{\max}) = \int_0^{t_{\max}} AUC(s, t) \cdot w(s) \, ds,
\]
where \(w(s)\) is a weight function.
The incident iAUC weights the AUC by a normalized function proportional to the product of the event time density \(f(t)\) and survival function \(S(t)\), which captures both the hazard and the number of subjects at risk. The integrated AUC is
\[
C_{\tau} = \int_0^{\tau} AUC(t) \cdot w_{\tau}(t) \, dt,
\]
where
\[
w_{\tau}(t) = \frac{2 \cdot f(t) \cdot S(t)}{W_{\tau}}, \quad \text{with} \quad W_{\tau} = \int_0^{\tau} 2 \cdot f(u) \cdot S(u) \, du.
\]
This incident iAUC corresponds to a global concordance measure related to Harrell's \(C\)-index for survival data \citep{heagerty2005survival}.\\
Similarly, the integrated Brier Score (iBS) summarizes overall prediction error over the time interval,
\[
iBS(t_{\max}) = \int_0^{t_{\max}} BS(s, t) w(s) ds,
\]
where \( w(s) \) denotes the chosen weighting scheme, such as those mentioned above.
For methodological and computational details, including adjustments for censoring and competing risks, see \citep{hashemi2025dynamic}, \citep{blanche2015quantifying}, and \citep{heagerty2005survival}.

\section{Simulation Study}
We evaluated the performance of the proposed zero-inflated joint model for longitudinal and survival data with a cure fraction through a comprehensive simulation study designed to emulate complex real-world data structures.
The simulation study comprises two main components:
\begin{itemize}
  \item {Estimation performance:} The first part assesses the estimation accuracy of our two proposed joint models with a cure component, based on two distinct approaches for modeling joint longitudinal and survival data in the presence of a cure fraction.
    \item {Dynamic prediction:} The second part evaluates the dynamic prediction performance of both models, allowing for a direct comparison of their ability to predict survival probabilities when a proportion of subjects are cured.
\end{itemize}
We explore a general scenario using two sets of the same real parameter values but with two different dispersion parameters. 
We then evaluate the performance of both the ZINBJMCF and ZIPJMCF.

\subsection{Data-Generating Mechanism}\label{this}
We simulated data for $N = 500$ and $N = 1000$ subjects, each with repeated measurements at observation times $s = \{0.0, 0.2, 0.4, \ldots, 2.0\}$ using the joint model $\text{ZINBJMCF}(\pi^\mathcal{Z}_i(s), \lambda_i(s), r)$.\\ 
For each subject \( i \), four baseline covariates were generated independently:  
\( x_{1i} \sim \text{Bernoulli}(0.5) \),  
\( x_{2i} \sim \mathcal{N}(0, 1) \),  
\( w_{1i} \sim \text{Bernoulli}(0.5)  \), and  
\( w_{2i} \sim \mathcal{N}(0, 1) \). 
The longitudinal count outcome \( Y_{i}(s) \), for subject \( i \) at time \( s \), follows a hurdle negative binomial model with:
\begin{equation} \label{zi2data}
\begin{aligned}
\eta_i^\lambda(s) &= \beta_{10} + \beta_{11} s + \beta_{12} x_{1i} + \beta_{13} x_{2i} + u_{i0} + u_{i1}s, \\
\eta_i^\mathcal{Z}(s) &= \beta_{20} + \beta_{21} s + \beta_{22} x_{1i} + b_{i0},
\end{aligned}
\end{equation}
where the parameter values are  
\( \bm{\beta}_1^\top =(\beta_{10},\beta_{11} , \beta_{12}, \beta_{13})= (0.5, 0.5, 0.5,-0.5) \),  
\( \bm{\beta}_2^\top =(\beta_{20},\beta_{21} , \beta_{22})= (-2, -0.5, 0.5)  \),  
and the dispersion parameter is \( r = 0.2 \) or $r=2$.
Additionally,  
\( (u_{i0}, u_{i1})^\top \sim \mathcal{N}_2(\bm{0}, \bm{D}_u) \) and  
\( b_{i0} \sim \mathcal{N}({0}, \sigma_b^2) \),  
where
\[
\bm{D}_u  = \begin{pmatrix} 1 & 0.5 \\ 0.5 & 1 \end{pmatrix}.
\] and $\sigma_b^2=1$
The probability of being cured was modeled using logistic regression:  
\[
\text{logit}(\pi^\mathcal{C}_i) = \xi_{10} + \xi_{11} w_{1i},
\]
with parameters \( \xi_{10} = -1 \), \( \xi_{11} = -1 \).\\
For susceptible subjects (\( \mathcal{C}_i = 1 \)), survival times were generated from a proportional hazards model with a constant baseline hazard \( h_0(t) = 1 \). The hazard function incorporated baseline covariates and the current value of the longitudinal process:
\[
h\left(t \mid \mathcal{C}_i = 1\right) = h_0(t) \exp\left(\xi_{21} w_{1i} + \xi_{22} w_{2i}+\alpha_1 \eta_i^\lambda(t) + \alpha_2 \eta_i^\mathcal{Z}(t)\right),
\]
where  \( \xi_{20} = -1 \), \( \xi_{21} = 1 \) and \( \alpha_1 = \alpha_2 = -0.5 \).\\
The survival data were simulated using a proportional hazards model and the inverse of the cumulative hazard function \citep{austin2012generating}, employing the inverse probability integral transform \citep{ross2012simulation}. This algorithm involves generating a random variable \( U \sim \text{Uniform}(0, 1) \), and solving the equation \( U = S(t) \) for \( t \).  
Given the constant baseline hazard, the survival function is defined as:  
\[
S_c(t) = \pi^\mathcal{C}_i + (1 - \pi^\mathcal{C}_i) S_i(t \mid \mathcal{C}_i = 1), \quad \text{with} \quad S_i(t \mid \mathcal{C}_i = 1) = \exp(-H(t_i|\mathcal{C}_i=1)),
\]
where the cumulative hazard function \( H(t_i) \) is expressed in closed form as:
\begin{equation} \label{chaz}
H(t_i) = \frac{h_0 \exp(A_{0i})}{A_{1i}} \left(\exp(A_{1i} t_i) - 1\right),
\end{equation}
with:
\begin{align*} \label{aoa1}
A_{0i} &= \xi_{21} w_{1i} + \xi_{22} w_{2i} + \alpha_1 \left(\beta_{10} + \beta_{12} x_{1i} + \beta_{13} x_{2i} + u_{i0}\right)  \\ &+ \alpha_2 \left(\beta_{20} + \beta_{22} x_{1i} + b_{i0}\right), \\
A_{1i} &= \alpha_1 \left(\beta_{11} + u_{i1}\right) + \alpha_2 \beta_{21}.
\end{align*}
To derive \( t_i \), we begin with the survival function:
\begin{align*}
U &= S_c(t_i) = \pi^\mathcal{C}_i + (1 - \pi^\mathcal{C}_i) \exp(-H(t_i)).
\end{align*}
Subtract \( \pi^\mathcal{C}_i \) from both sides and divide by \( (1 - \pi^\mathcal{C}_i) \):
\[
\exp(-H(t_i)) = \frac{U - \pi^\mathcal{C}_i}{1 - \pi^\mathcal{C}_i}.
\]
Taking the natural logarithm of both sides:
\[
-H(t_i) = \log\left( \frac{U - \pi^\mathcal{C}_i}{1 - \pi^\mathcal{C}_i} \right).
\]
Finally, solving for \( t_i \) using Equation~\eqref{chaz} yields:
\begin{align}
t_i = \frac{1}{A_{1i}} \log\left[ 1 - \frac{A_{1i}}{h_0 \exp(A_{0i})} \log\left( \frac{U - \pi^\mathcal{C}_i}{1 - \pi^\mathcal{C}_i} \right) \right]. \label{ti}
\end{align}
If $U - \pi^\mathcal{C}_i < 0$, subject $i$ was considered censored and classified as not cured at the follow-up time.

\subsection{Assessing Parameter Estimation Accuracy}
To evaluate parameter estimation, we fitted the joint model $\text{ZINBJMCF}(\pi^\mathcal{Z}_i(s), \lambda_i(s), r)$. Estimation accuracy was quantified using relative bias (RB), root mean square error (RMSE), and coverage rate (CR) for all parameters. For Bayesian inference, two parallel Hamiltonian Monte Carlo (HMC) chains were run for 2,000 iterations each; the first 1,000 iterations of each chain were discarded as burn-in (warmup), and the remaining 1,000 iterations were used to compute posterior summaries. For comparison, we also evaluated the joint model $\text{ZIPJMCF}(\pi^\mathcal{Z}_i(s), \lambda_i(s))$. Although in these simulation studies we used only HMC, the results can also be obtained using VB with the  R code available at \url{https://github.com/tbaghfalaki/HJMCF/tree/main/Simulation} using the \texttt{rstan} R package.\\ The simulation results, for $r=0.2$ and $r=2$ presented in Tables \ref{r01} and \ref{r02}, respectively.   When the data were generated from the ZINB joint model, the correctly specified ZINBJM produced nearly unbiased estimates for most parameters in both scenarios ($r = 0.2$ and $r = 2$), with RMSE values close to the empirical SDs and coverage rates generally near the nominal 95\%, especially for larger sample sizes. In contrast, the misspecified ZIPJM showed substantial bias and poor coverage for several key parameters, particularly when overdispersion was high ($r = 0.2$), with extreme bias observed for the intercept $\beta_{10}$, the random-effect variances ($D_{u,11}$, $D_{u,22}$), and the correlation  $\rho_u$ (between $u_{i0}$ and $u_{i1}$), often yielding coverage rates close to zero. Although increasing the sample size reduced variability for both models, it did not mitigate the systematic bias in ZIPJM estimates, underscoring the importance of correctly modeling overdispersion in zero-inflated longitudinal--survival data.

\begin{landscape}
\begin{table}[h] 
 \caption{\label{r01} Results of the simulation study for data generated from a Zero-Inflated Negative Binomial (ZINB) joint model with $r = 0.2$ and survival time including a cure fraction. Analyses were performed under both the ZINB joint model (ZINBJMCF) and the Zero-Inflated Poisson joint model (ZIPJMCF). For each parameter, the table reports the mean estimate (Est.), standard deviation (SD), relative bias (RB), root mean square error (RMSE), and coverage rate (CR), based on 100 simulated datasets with sample sizes of 500 and 1000.}
\centering
\tiny
\begin{tabular}{c|c|ccccc|ccccc}
  \hline
 & &\multicolumn{5}{c|}{N=500} &\multicolumn{5}{c}{N=1000} \\    
 \hline
 & & \multicolumn{9}{c}{ZINBJMCF} \\    
 \hline
Parameter & Real & Est. & SD & RB & RMSE & CR & Est. & SD & RB & RMSE & CR\\ 
  \hline
$\beta_{10}$ & 0.500 & 0.386 & 0.098 & -0.228 & 0.115 & 94 & 0.465 & 0.098 & -0.070 & 0.110 & 96 \\
$\beta_{11}$ & 0.500 & 0.532 & 0.100 & 0.064 & 0.103 & 98 & 0.528 & 0.090 & 0.056 & 0.103 & 95 \\
$\beta_{12}$ & 0.500 & 0.590 & 0.140 & 0.180 & 0.208 & 91 & 0.513 & 0.122 & 0.026 & 0.125 & 95 \\
$\beta_{13}$ & -0.500 & -0.491 & 0.073 & -0.018 & 0.073 & 92 & -0.504 & 0.059 & 0.008 & 0.059 & 97 \\
$\beta_{20}$ & -2.000 & -2.029 & 0.098 & 0.015 & 0.101 & 97 & -2.017 & 0.078 & 0.008 & 0.079 & 98 \\
$\beta_{21}$ & -0.500 & -0.488 & 0.065 & -0.024 & 0.065 & 95 & -0.499 & 0.061 & -0.001 & 0.061 & 96 \\
$\beta_{22}$ & 0.500 & 0.463 & 0.108 & -0.074 & 0.128 & 96 & 0.487 & 0.074 & -0.026 & 0.082 & 95 \\
$\zeta_{10}$ & -1.000 & -1.052 & 0.192 & 0.052 & 0.200 & 92 & -1.004 & 0.186 & 0.004 & 0.187 & 96 \\
$\zeta_{11}$ & -1.000 & -1.285 & 0.476 & 0.285 & 0.797 & 97 & -1.013 & 0.173 & 0.013 & 0.174 & 93 \\
$\zeta_{20}$ & -0.500 & -0.490 & 0.086 & -0.020 & 0.090 & 97 & -0.489 & 0.069 & -0.022 & 0.087 & 97 \\
$\zeta_{21}$ & -0.500 & -0.522 & 0.197 & 0.044 & 0.202 & 92 & -0.485 & 0.112 & -0.030 & 0.118 & 98 \\
$\alpha_1$ & -1.000 & -1.025 & 0.185 & 0.025 & 0.191 & 95 & -1.021 & 0.158 & 0.021 & 0.162 & 95 \\
$\alpha_2$ & 1.000 & 0.973 & 0.087 & -0.027 & 0.092 & 92 & 0.980 & 0.075 & -0.020 & 0.078 & 96 \\
$r$ & 0.200 & 0.195 & 0.022 & -0.025 & 0.029 & 94 & 0.196 & 0.022 & -0.020 & 0.030 & 94 \\
$D_{u,11}$ & 1.000 & 1.001 & 0.045 & 0.001 & 0.045 & 96 & 0.975 & 0.049 & -0.025 & 0.050 & 98 \\
$D_{u,22}$ & 1.000 & 1.021 & 0.109 & 0.021 & 0.111 & 94 & 1.030 & 0.068 & 0.030 & 0.072 & 98 \\
$\sigma_b^2$ & 1.000 & 1.032 & 0.114 & 0.032 & 0.125 & 91 & 1.039 & 0.094 & 0.039 & 0.104 & 96 \\
$D_{u,12}$ & 0.500 & 0.467 & 0.184 & -0.066 & 0.196 & 97 & 0.478 & 0.146 & -0.045 & 0.152 & 96 \\
\hline
 & & \multicolumn{9}{c}{ZIPJMCF} \\    
 \hline
$\beta_{10}$ & 0.500 & 1.092 & 0.145 & 1.184 & 1.103 & 0 & 1.090 & 0.108 & 1.180 & 1.095 & 0 \\
$\beta_{11}$ & 0.500 & 0.513 & 0.144 & 0.025 & 0.146 & 28 & 0.526 & 0.105 & 0.052 & 0.115 & 19 \\
$\beta_{12}$ & 0.500 & 0.430 & 0.185 & -0.140 & 0.198 & 44 & 0.393 & 0.134 & -0.214 & 0.173 & 29 \\
$\beta_{13}$ & -0.500 & -0.391 & 0.082 & -0.218 & 0.159 & 21 & -0.383 & 0.065 & -0.233 & 0.145 & 6 \\
$\beta_{20}$ & -2.000 & -2.029 & 0.146 & 0.015 & 0.150 & 86 & -2.019 & 0.114 & 0.010 & 0.116 & 77 \\
$\beta_{21}$ & -0.500 & -0.488 & 0.095 & -0.024 & 0.104 & 92 & -0.497 & 0.071 & -0.006 & 0.073 & 89 \\
$\beta_{22}$ & 0.500 & 0.508 & 0.156 & 0.016 & 0.156 & 84 & 0.500 & 0.112 & 0.000 & 0.112 & 86 \\
$\zeta_{10}$ & -1.000 & -1.045 & 0.206 & 0.045 & 0.214 & 88 & -1.028 & 0.149 & 0.028 & 0.154 & 88 \\
$\zeta_{11}$ & -1.000 & -2.377 & 0.677 & 1.377 & 2.271 & 72 & -1.395 & 0.023 & 0.395 & 2.049 & 66 \\
$\zeta_{20}$ & -0.500 & -0.442 & 0.080 & -0.116 & 0.133 & 65 & -0.440 & 0.061 & -0.119 & 0.083 & 49 \\
$\zeta_{21}$ & -0.500 & -0.558 & 0.214 & 0.116 & 0.253 & 61 & -0.544 & 0.150 & 0.088 & 0.206 & 52 \\
$\alpha_1$ & -1.000 & -0.951 & 0.211 & -0.049 & 0.219 & 80 & -0.958 & 0.160 & -0.042 & 0.167 & 77 \\
$\alpha_2$ & 1.000 & 0.948 & 0.111 & -0.052 & 0.123 & 73 & 0.952 & 0.070 & -0.048 & 0.086 & 77 \\
$D_{u,11}$ & 1.000 & 1.534 & 0.160 & 0.534 & 0.549 & 0 & 1.538 & 0.117 & 0.538 & 0.550 & 0 \\
$D_{u,22}$ & 1.000 & 1.836 & 0.219 & 0.836 & 0.853 & 0 & 1.835 & 0.148 & 0.835 & 0.847 & 0 \\
$\sigma_b^2$ & 1.000 & 0.990 & 0.115 & -0.010 & 0.115 & 58 & 0.992 & 0.086 & -0.008 & 0.086 & 54 \\
$D_{u,12}$ & 0.500 & -0.426 & 0.085 & -1.852 & 0.932 & 0 & -0.431 & 0.066 & -1.862 & 0.934 & 0 \\
\hline
\end{tabular}
\vspace*{6pt}
\end{table} 
\end{landscape}
\FloatBarrier

\begin{landscape}
\begin{table}[h] 
 \caption{\label{r02} Results of the simulation study for data generated from a Zero-Inflated Negative Binomial (ZINB) joint model with $r = 2$ and survival time including a cure fraction. Analyses were performed under both the ZINB joint model (ZINBJMCF) and the Zero-Inflated Poisson joint model (ZIPJMCF). For each parameter, the table reports the mean estimate (Est.), standard deviation (SD), relative bias (RB), root mean square error (RMSE), and coverage rate (CR), based on 100 simulated datasets with sample sizes of 500 and 1000.}
\centering
\tiny
\begin{tabular}{c|c|ccccc|ccccc}
  \hline
 & &\multicolumn{5}{c|}{N=500} &\multicolumn{5}{c}{N=1000} \\    
 \hline
 & & \multicolumn{9}{c}{ZINBJMCF} \\    
 \hline
Parameter & Real & Est. & SD & RB & RMSE & CR & Est. & SD & RB & RMSE & CR\\ 
  \hline
$\beta_{10}$ & 0.500 & 0.566 & 0.100 & 0.132 & 0.114 & 93 & 0.503 & 0.059 & 0.006 & 0.059 & 98 \\
$\beta_{11}$ & 0.500 & 0.535 & 0.083 & 0.070 & 0.088 & 94 & 0.512 & 0.034 & 0.024 & 0.036 & 99 \\
$\beta_{12}$ & 0.500 & 0.589 & 0.122 & 0.178 & 0.148 & 96 & 0.537 & 0.104 & 0.074 & 0.107 & 93 \\
$\beta_{13}$ & -0.500 & -0.507 & 0.069 & 0.013 & 0.069 & 95 & -0.501 & 0.040 & -0.002 & 0.040 & 93 \\
$\beta_{20}$ & -2.000 & -2.031 & 0.103 & 0.016 & 0.107 & 94 & -2.034 & 0.097 & 0.017 & 0.099 & 95 \\
$\beta_{21}$ & -0.500 & -0.487 & 0.072 & -0.026 & 0.072 & 94 & -0.501 & 0.068 & 0.002 & 0.068 & 98 \\
$\beta_{22}$ & 0.500 & 0.464 & 0.106 & -0.072 & 0.112 & 94 & 0.488 & 0.093 & -0.024 & 0.096 & 98 \\
$\zeta_{10}$ & -1.000 & -1.016 & 0.203 & 0.016 & 0.204 & 91 & -1.001 & 0.220 & 0.001 & 0.220 & 93 \\
$\zeta_{11}$ & -1.000 & -1.338 & 0.464 & 0.338 & 0.591 & 94 & -1.158 & 0.226 & 0.158 & 0.271 & 92 \\
$\zeta_{20}$ & -0.500 & -0.497 & 0.074 & -0.006 & 0.074 & 93 & -0.477 & 0.050 & -0.045 & 0.067 & 97 \\
$\zeta_{21}$ & -0.500 & -0.549 & 0.161 & 0.098 & 0.174 & 92 & -0.483 & 0.116 & -0.035 & 0.122 & 97 \\
$\alpha_1$ & -1.000 & -0.969 & 0.193 & -0.031 & 0.196 & 93 & -0.950 & 0.115 & -0.050 & 0.124 & 96 \\
$\alpha_2$ & 1.000 & 0.994 & 0.086 & -0.006 & 0.086 & 93 & 0.944 & 0.043 & -0.056 & 0.068 & 98 \\
$r$ & 2.000 & 1.860 & 0.484 & -0.070 & 0.504 & 94 & 1.945 & 0.105 & -0.028 & 0.111 & 83 \\
$D_{u,11}$ & 1.000 & 1.001 & 0.058 & 0.001 & 0.058 & 96 & 0.964 & 0.048 & -0.036 & 0.059 & 95 \\
$D_{u,22}$ & 1.000 & 1.010 & 0.072 & 0.010 & 0.073 & 96 & 1.028 & 0.042 & 0.028 & 0.050 & 99 \\
$\sigma_b^2$ & 1.000 & 1.002 & 0.122 & 0.002 & 0.122 & 97 & 1.008 & 0.061 & 0.008 & 0.061 & 93 \\
$D_{u,12}$ & 0.500 & 0.491 & 0.117 & -0.018 & 0.118 & 97 & 0.486 & 0.069 & -0.028 & 0.073 & 96 \\
\hline
 & & \multicolumn{9}{c}{ZIPJMCF} \\    
 \hline
$\beta_{10}$ & 0.500 & 0.583 & 0.119 & 0.166 & 0.146 & 48 & 0.578 & 0.078 & 0.157 & 0.092 & 38 \\
$\beta_{11}$ & 0.500 & 0.507 & 0.108 & 0.014 & 0.108 & 45 & 0.504 & 0.067 & 0.007 & 0.067 & 26 \\
$\beta_{12}$ & 0.500 & 0.492 & 0.140 & -0.016 & 0.140 & 51 & 0.470 & 0.106 & -0.060 & 0.122 & 49 \\
$\beta_{13}$ & -0.500 & -0.454 & 0.068 & -0.092 & 0.081 & 46 & -0.459 & 0.048 & -0.081 & 0.061 & 40 \\
$\beta_{20}$ & -2.000 & -2.003 & 0.138 & 0.001 & 0.138 & 84 & -2.005 & 0.101 & 0.003 & 0.101 & 84 \\
$\beta_{21}$ & -0.500 & -0.488 & 0.096 & -0.023 & 0.096 & 90 & -0.495 & 0.075 & -0.010 & 0.075 & 84 \\
$\beta_{22}$ & 0.500 & 0.483 & 0.148 & -0.035 & 0.149 & 89 & 0.485 & 0.110 & -0.031 & 0.111 & 83 \\
$\zeta_{10}$ & -1.000 & -1.039 & 0.191 & 0.039 & 0.195 & 88 & -1.025 & 0.145 & 0.025 & 0.147 & 85 \\
$\zeta_{11}$ & -1.000 & -3.251 & 10.407 & 2.251 & 10.730 & 72 & -1.119 & 0.492 & 0.119 & 0.503 & 65 \\
$\zeta_{20}$ & -0.500 & -0.515 & 0.076 & 0.030 & 0.076 & 83 & -0.509 & 0.054 & 0.019 & 0.054 & 85 \\
$\zeta_{21}$ & -0.500 & -0.569 & 0.195 & 0.138 & 0.201 & 65 & -0.535 & 0.142 & 0.069 & 0.148 & 46 \\
$\alpha_1$ & -1.000 & -1.018 & 0.201 & 0.018 & 0.201 & 78 & -1.000 & 0.150 & 0.000 & 0.150 & 76 \\
$\alpha_2$ & 1.000 & 1.005 & 0.113 & 0.005 & 0.113 & 84 & 1.000 & 0.081 & 0.000 & 0.081 & 79 \\
$D_{u,11}$ & 1.000 & 1.091 & 0.055 & 0.091 & 0.106 & 43 & 1.096 & 0.039 & 0.096 & 0.103 & 11 \\
$D_{u,22}$ & 1.000 & 1.163 & 0.080 & 0.163 & 0.182 & 19 & 1.168 & 0.065 & 0.168 & 0.179 & 4 \\
$\sigma_b^2$ & 1.000 & 0.976 & 0.101 & -0.024 & 0.103 & 75 & 0.991 & 0.084 & -0.009 & 0.084 & 51 \\
$D_{u,12}$ & 0.500 & 0.092 & 0.107 & -0.816 & 0.423 & 0 & 0.093 & 0.071 & -0.815 & 0.413 & 0 \\
\hline
\end{tabular}
\vspace*{6pt}
\end{table} 
\end{landscape}
\FloatBarrier

\subsection{Assessing Dynamic Risk Prediction Performance}
To evaluate the risk prediction performance of zero-inflated joint models with a cure fraction, we generated data as described in Section~\ref{this} with a sample size of \(N = 1000\). The dataset was evenly split into training (50\%) and validation (50\%) sets. Parameters were estimated using the training data, while risk predictions were evaluated on the validation set at landmark times \(s = 0, 0.5,\) and \(1\), with a prediction window of $0.5$. We compared the predictive performance of the ZINBJMCF and the ZIPJMCF. Following Section~\ref{Riskaids}, evaluation metrics included the AUC, BS, iAUC, and iBS. For reference, real values of dynamic probabilities were calculated using the true parameter values and the initially generated random effects, which remain the same across different dispersion scenarios. The results, summarized in Table~\ref{predee}, indicate that both models achieve reasonable predictive accuracy across scenarios.  \\
For both dispersion settings, ZINBJMCF and ZIPJMCF showed similar predictive performance. In the low-dispersion case (\(r=0.2\)), AUC values were moderately high (around 0.62--0.74), indicating reasonable discrimination, though slightly below the true AUC values (0.71--0.76). BS values were low (0.11--0.19), suggesting acceptable prediction accuracy, but slightly overestimated compared to the real BS (0.10--0.15). In the high-dispersion case (\(r=2\)), AUC estimates improved slightly, moving closer to the true values, while BS values remained consistently low, indicating stable calibration. Overall, both models slightly underestimate discrimination (AUC) and slightly overestimate prediction error (BS), but maintain robust predictive performance across different dispersion levels, confirming reliability for dynamic risk prediction in the presence of zero inflation and a cure fraction.

\begin{table}[h!]
\caption{\label{predee}Estimates (Est) and Standard Deviations (SD) of AUC and Brier Score (BS) over time for ZINBJMCF and ZIPJMCF models, including Integrated metrics}
\centering
\begin{tabular}{cc|cc|cc|cc}
\hline   
&  & \multicolumn{6}{c}{$r=0.2$} \\\hline
&  & \multicolumn{2}{c|}{ZINBJMCF} & \multicolumn{2}{c|}{ZIPJMCF} & \multicolumn{2}{c}{Real} \\\hline
& Landmark Time  & Est & SD & Est & SD & Est & SD \\
\hline
   & 0   & 0.731 & 0.031 & 0.741 & 0.025 & 0.756 & 0.030 \\
AUC& 0.5 & 0.635 & 0.041 & 0.639 & 0.045 & 0.714 & 0.025 \\
   & 1   & 0.618 & 0.053 & 0.596 & 0.052 & 0.706 & 0.036 \\
\hline
   & 0   & 0.187 & 0.011 & 0.193 & 0.008 & 0.173 & 0.015 \\
BS & 0.5 & 0.154 & 0.012 & 0.153 & 0.011 & 0.144 & 0.014 \\
   & 1   & 0.111 & 0.011 & 0.107 & 0.011 & 0.100 & 0.011 \\
\hline
\multicolumn{2}{c|}{iAUC}   & 0.699 & 0.023 & 0.698 & 0.024 & 0.738 & 0.023 \\
\multicolumn{2}{c|}{iBS} & 0.169 & 0.010 & 0.172 & 0.008 & 0.156 & 0.013 \\  
\hline
&  & \multicolumn{6}{c}{$r=2$} \\\hline
   & 0   & 0.733 & 0.032 & 0.735 & 0.021 & 0.756 & 0.030 \\
AUC& 0.5 & 0.638 & 0.045 & 0.627 & 0.041 & 0.714 & 0.025 \\
   & 1   & 0.634 & 0.060 & 0.628 & 0.065 & 0.706 & 0.036 \\
\hline
   & 0   & 0.187 & 0.011 & 0.195 & 0.007 & 0.173 & 0.015 \\
BS & 0.5 & 0.153 & 0.011 & 0.154 & 0.011 & 0.144 & 0.014 \\
   & 1   & 0.108 & 0.009 & 0.108 & 0.012 & 0.100 & 0.011 \\
\hline
\multicolumn{2}{c|}{iAUC}   & 0.702 & 0.026 & 0.699 & 0.020 & 0.738 & 0.023 \\
\multicolumn{2}{c|}{iBS} & 0.168 & 0.010 & 0.174 & 0.007 & 0.156 & 0.013 \\
\hline
\end{tabular}
\end{table}

\section{Application: AIDS data}
\subsection{Data Description}
\begin{figure}[hbt!]
\centering
\includegraphics[width=9cm]{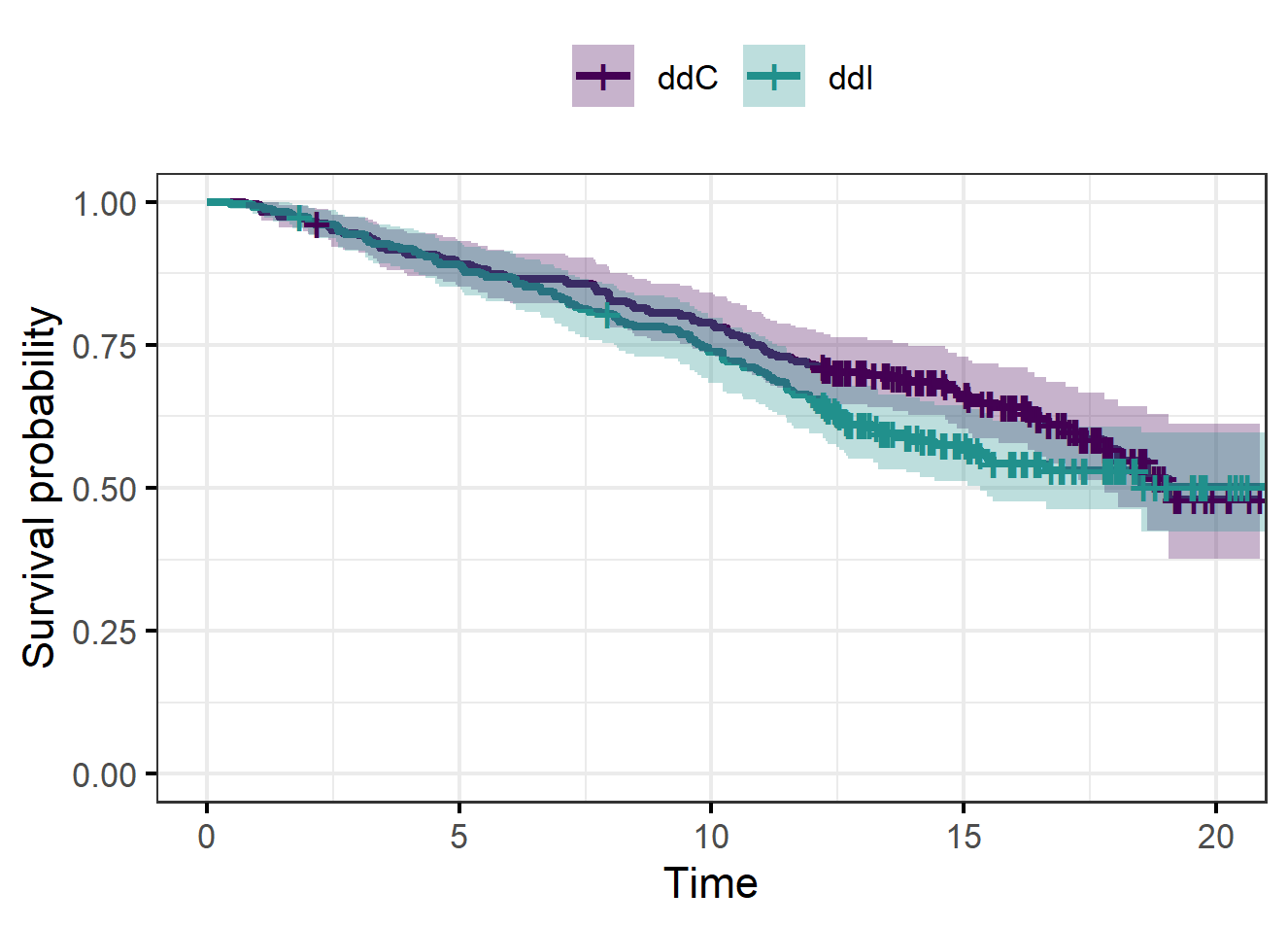}
\vspace*{-.1cm} \caption{\label{surhiv}
Kaplan-Meier survival curve for time to death, along with its $95\%$ confidence bands, for HIV data.}
\end{figure}

In this section, we reanalyze a dataset from a longitudinal study of HIV-infected individuals, comprising 467 patients ($N = 467$). The study includes repeated measurements of CD4 cell counts, a type of white blood cell essential to immune function. CD4 cells play a key role in combating infections by activating the immune system against viruses, bacteria, and other pathogens.
\\
Numerous studies have examined the joint modeling of longitudinal CD4 measurements and time-to-event outcomes, often using the square root of CD4 as a continuous marker. More recently, CD4 counts have been modeled as zero-inflated count data, as in \citep{baghfalaki2021approximate, ganjali2024joint}. However, these studies focused solely on the survival component and did not account for the cure fraction in mortality. To assess the suitability of a zero-inflated model for this dataset, as discussed in \citep{baghfalaki2021approximate, ganjali2024joint}, the test of \citep{friendly2017package} strongly rejected the null hypothesis of no zero-inflation, reporting a $\mathcal{X}^2$ statistic with a p-value of 0.000. This provides strong evidence in favor of zero inflation in the data.
\\
In this data, participants were randomly assigned to receive one of two antiretroviral drugs: didanosine (ddI) or zalcitabine (ddC). CD4 cell counts were measured at baseline (time of randomization) and at 2, 6, 12, and 18 months. These repeated CD4 measurements serve as the longitudinal outcome variable.
\\
Covariates in the analysis include gender, treatment group (ddI vs. ddC), previous opportunistic infection (PrevOI: AIDS diagnosis vs. no AIDS diagnosis), and AZT treatment status (failure vs. intolerance), where AZT refers to zidovudine, an early antiretroviral drug.
\\
Time to death is considered the event of interest. The Kaplan--Meier survival curves by treatment group, presented in Figure~\ref{surhiv}, exhibit a characteristic plateau toward the end of the follow-up period, suggesting the presence of a subpopulation of long-term survivors.

\subsection{Data Analysis}\label{data_analysis}
The objective of this study is to jointly model the longitudinal trajectory of CD4 counts and the time-to-event outcome with a cure fraction, while accounting for a potential excess of zero values in the CD4 measurements.
\\
To describe the longitudinal process, we specify two linear predictors: one for the rate (intensity) of CD4 counts, and another for the probability of observing a zero count at each time point \( s_{ij} \). These predictors are defined as follows:
\begin{align}\label{lamaids}
\eta_i^\lambda (s_{ij}) = \log (\lambda (s_{ij})) &= \beta_{10} + \beta_{11}s_{ij} + \beta_{12}\text{Gender}_i + \beta_{13}\text{Drug}_i + \beta_{14}\text{PrevOI}_i \\\nonumber
&\quad + \beta_{15}\text{AZT}_i + u_{i0} + u_{i1}s_{ij}, 
\end{align}
\begin{align}\label{piaids}
\eta_i^\mathcal{Z} (s_{ij}) = \text{logit}(\pi (s_{ij})) &= \beta_{20} + \beta_{21}s_{ij} + \beta_{22}\text{PrevOI}_i + b_{i0},
\end{align}
where the subject-specific random effects follow bivariate and univariate normal distributions as
$(u_{i0}, u_{i1})^\top \sim \mathcal{N}_2(\mathbf{0}, \mathbf{D}_u)$ and $b_{i0} ^\top \sim \mathcal{N}({0}, \sigma^2_b)$, respectively.\\
The time-to-death sub-model depends on the current values of the two longitudinal components \( \eta_i^\lambda(t) \) and \( \eta_i^\mathcal{Z}(t) \), and includes adjustments for baseline covariates. The hazard function is defined as:
\[
h(t \mid \mathcal{C}_i = 1) = h_0(t) \exp\left\{ \alpha_1  \eta_i^\lambda (t) + \alpha_2  \eta_i^\mathcal{Z} (t) \right\}.
\]
Additionally, the probability of being cured is modeled using the following logistic regression:
\begin{align}\label{picureaids}
\text{logit}(\pi_i^{\mathcal{C}}) = \xi_{10} + 
\xi_{11}\text{Gender}_i + \xi_{12}\text{Drug}_i + \xi_{13}\text{PrevOI}_i + \xi_{14}\text{AZT}_i
\end{align}
The results of this model is given in Table \ref{appli}, which provides insights into the effects of covariates on the longitudinal CD4 counts, the probability of observing zero counts, survival, and the cure fraction in the joint ZINB model and the ZIP model. 
Parameter estimation was carried out using HMC implemented in the \texttt{rstan} package, with 5000 iterations, the first 2500 serving as warmup (burn-in). Convergence of the parameters was evaluated using the potential scale reduction factor $\hat{R}$ (Brooks--Gelman--Rubin diagnostic) \citep{gelman1992inference}. The R code for parameter estimation and dynamic prediction is available at \url{https://github.com/tbaghfalaki/HJMCF/tree/main/AIDS}.
First of all, model comparison metrics (WAIC and BIC) indicate that the ZINB model provides a better fit than that of the ZIP model, reflecting the importance of accommodating overdispersion in the longitudinal CD4 counts.
\\
In the count sub-model, time has a significant negative effect on HIV counts (Est. = --0.058, 95\% CI: --0.070, --0.046), indicating a gradual decline over follow-up. A prior AIDS diagnosis (PrevOI) is strongly associated with lower counts (Est. = --1.214, 95\% CI: --1.559, --0.856), while gender, treatment with ddI, and AZT failure do not show statistically significant effects. The estimated dispersion parameter ($r = 5.304$) in the ZINB model confirms the presence of overdispersion compared to the Poisson model.  
\\
In the zero-inflation sub-model, PrevOI has a strong positive association with the probability of an excess zero (Est. = 2.769, 95\% CI: 1.085, 5.622), suggesting that individuals with a prior AIDS diagnosis are more likely to have structural zeros, while time shows no significant effect.  
\\
In the survival sub-model, the zero-inflation probability ($\alpha_2$) is positively associated with the hazard of death (Est. = 0.399, 95\% CI: 0.035, 0.890), implying that patients more prone to structural zeros have higher mortality risk. The current HIV count level ($\alpha_1$) is not significantly related to hazard. The baseline hazard parameters ($h^*_1, \dots, h^*_4$) capture time-varying risk patterns but have wide credible intervals.  
\\
In the cure fraction sub-model, none of the regression coefficients ($\xi_{10} \dots \xi_{14}$) are statistically significant, indicating that the included covariates do not strongly predict membership in the cured fraction. The posterior probability that subject $i$ belongs to the cured fraction, given that they have survived beyond their observed time $t_i$, is
\[
\text{P}^\text{cure}_{i}=P(\text{cured} \mid T_i > t_i) = \frac{P(\text{cured and } T_i > t_i)}{P(T_i > t_i)} = \frac{\pi_i^{\mathcal{C}}}{\pi_i^{\mathcal{C}} + (1 - \pi_i^{\mathcal{C}})S(t_i|\mathcal{C}_i=1)},
\]
where $\pi_i^{\mathcal{C}}$ is the prior cure probability and $S(t_i)$ is the survival function for uncured subjects. The population-level average posterior cure probability is then
\[
\bar{\text{P}}^\text{cure}= \frac{1}{N} \sum_{i=1}^{N} \text{P}^\text{cure}_{i},
\]
representing the expected fraction of patients in the cured group across the study population. Across all subjects, the posterior mean is $\bar{\text{P}}^\text{cure}= 0.192$~ (95\% \text{CI}: 0.035, 0.391), indicating that approximately 19\% of patients are estimated to be long-term survivors (“cured”), with some uncertainty.

 \begin{table}[h]
 \caption{\label{appli} Parameter estimates (Est.), standard deviation (SD), 2.5\% and 97.5\% quantiles of the 95\% credible interval, and $\hat{R}$ (Gelman-Rubin diagnostic) for analyzing HIV data using zero-inflated negative binomial (ZINB) joint models with a mixture cure component.}
\centering
\tiny
\begin{tabular}{c|ccccc|ccccc}
\hline    
Parameter & \multicolumn{5}{c|}{ZINBJMCF} & \multicolumn{5}{c}{ZIPJMCF } \\\hline
                           & Est. & SD & 2.5\% & 97.5\% & $\hat{R}$ & Est. & SD & 2.5\% & 97.5\% & $\hat{R}$ \\
\hline
\multicolumn{11}{c}{{Count Model}} \\\hline
Intercept       &  4.176 & 0.254 &  3.645 &  4.673 & 1.000 &  4.100 & 0.260 &  3.606 &  4.597 & 1.016 \\
Time      & -0.058 & 0.006 & -0.070 & -0.046 & 1.000 & -0.064 & 0.009 & -0.081 & -0.048 & 1.036 \\
Gender (Male)      &  0.284 & 0.241 & -0.176 &  0.791 & 1.000 &  0.293 & 0.252 & -0.189 &  0.788 & 1.005 \\
Drug (ddI)      & -0.011 & 0.134 & -0.266 &  0.248 & 1.000 & -0.021 & 0.144 & -0.294 &  0.283 & 1.000 \\
PrevOI (AIDS)      & -1.214 & 0.180 & -1.559 & -0.856 & 1.000 & -1.209 & 0.181 & -1.556 & -0.874 & 1.029 \\
AZT (failure)      & -0.136 & 0.172 & -0.471 &  0.182 & 1.001 & -0.141 & 0.177 & -0.473 &  0.217 & 1.000 \\
$d_{u,11}$ (var)   &  1.165 & 0.054 &  1.063 &  1.276 & 1.000 &  1.226 & 0.050 &  1.131 &  1.330 & 1.000 \\
$d_{u,22}$ (var)   &  0.057 & 0.007 &  0.042 &  0.071 & 1.000 &  0.130 & 0.008 &  0.115 &  0.147 & 1.000 \\
$\rho_u$ (corr)   &  0.185 & 0.122 & -0.057 &  0.426 & 1.007 & -0.063 & 0.071 & -0.200 &  0.079 & 1.002 \\
Dispersion ($r$)          &  5.304 & 0.461 &  4.434 &  6.238 & 1.001 &   --   &   --  &   --   &   --   &   --  \\
\hline
\multicolumn{11}{c}{{Zero-Inflated Model}} \\\hline
Intercept      & -6.813 & 1.259 & -9.898 & -4.972 & 1.018 & -6.583 & 1.123 & -9.193 & -4.865 & 1.001 \\
Time     &  0.035 & 0.051 & -0.068 &  0.133 & 1.000 &  0.032 & 0.053 & -0.075 &  0.128 & 1.000 \\
PrevOI (AIDS)      &  2.769 & 1.152 &  1.085 &  5.622 & 1.005 &  2.622 & 1.035 &  1.065 &  5.120 & 1.000 \\
$\sigma^2_{b}$ (var)       &  0.656 & 0.351 &  0.166 &  1.576 & 1.113 &  0.511 & 0.337 &  0.085 &  1.309 & 1.039 \\\hline
\multicolumn{11}{c}{{Cure Fraction Model}} \\\hline
$\xi_{10}$  & -0.303 & 0.887 & -2.128 &  1.303 & 1.001 & -0.345 & 0.883 & -2.232 &  1.277 & 1.001 \\
$\xi_{11}$ &  0.009 & 0.753 & -1.577 &  1.373 & 1.003 & -0.059 & 0.784 & -1.713 &  1.372 & 1.002 \\
$\xi_{12}$ & -0.914 & 0.675 & -2.335 &  0.302 & 1.000 & -0.917 & 0.677 & -2.375 &  0.330 & 1.000 \\
$\xi_{13}$ & -1.299 & 0.803 & -2.819 &  0.332 & 1.001 & -1.280 & 0.788 & -2.810 &  0.312 & 1.002 \\
$\xi_{14}$ & -0.897 & 0.723 & -2.354 &  0.500 & 1.000 & -0.882 & 0.763 & -2.437 &  0.593 & 1.000 \\
\hline
\multicolumn{11}{c}{{Survival Model}} \\\hline
$\alpha_1$        &  0.024 & 0.089 & -0.152 &  0.204 & 1.000 &  0.001 & 0.078 & -0.153 &  0.155 & 1.000 \\
$\alpha_2$        &  0.399 & 0.222 & 0.035 &  0.890 & 1.012 &  0.434 & 0.235 &  0.008 &  0.926 & 1.003 \\
$h^*_1$          &  0.280 & 0.325 &  0.034 &  1.203 & 1.005 &  0.351 & 0.453 &  0.042 &  1.464 & 1.001 \\
$h^*_2$          &  0.553 & 0.657 &  0.070 &  2.385 & 1.003 &  0.684 & 0.900 &  0.081 &  2.930 & 1.001 \\
$h^*_3$         &  0.499 & 0.630 &  0.060 &  2.131 & 1.003 &  0.606 & 0.793 &  0.071 &  2.680 & 1.001 \\
$h^*_4$          &  0.809 & 1.056 &  0.091 &  3.870 & 1.001 &  0.972 & 1.309 &  0.106 &  4.428 & 1.002 \\
\hline
  WAIC            & \multicolumn{5}{c|}{ 9563.0} & \multicolumn{5}{c}{ 13158.3} \\  
  BIC             & \multicolumn{5}{c|}{ 9333.43} & \multicolumn{5}{c}{ 10885.22} \\  
  \hline
\end{tabular}
\end{table}
\FloatBarrier

\subsection{Risk predictions}\label{Riskaids}
\begin{figure}[hbt!]
\centering
\includegraphics[width=11cm]{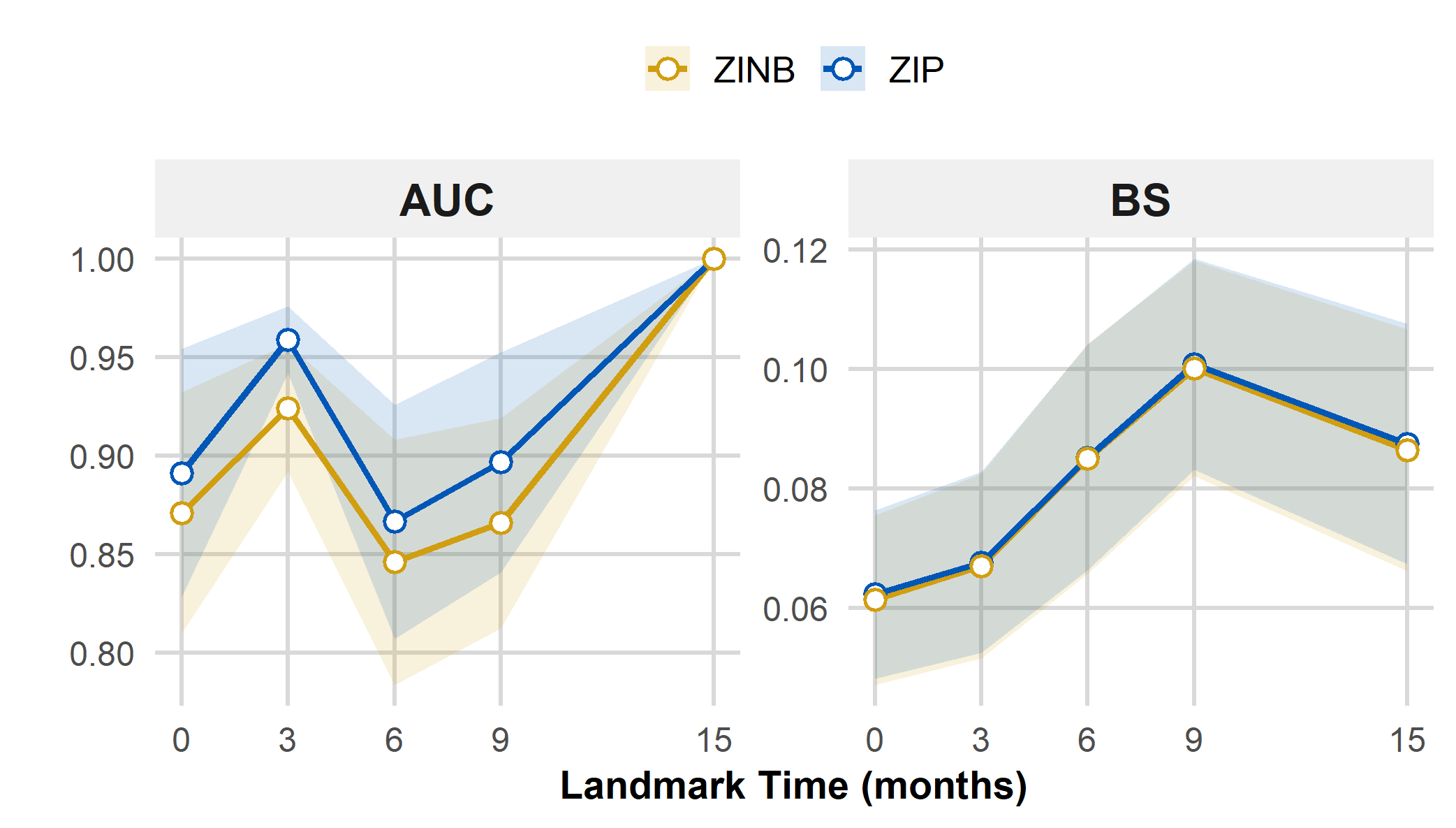}
\vspace*{-.1cm} \caption{\label{auc1}
Time-dependent predictive performance of ZINBJMCF and ZIPJMCF models, evaluated at landmark times 0, 3, 6, 9, and 15 months using a three-month prediction window, based on the area under the ROC curve (AUC) and the Brier score (BS).}
\end{figure}

For risk prediction, we used 70\% of the data as the training set and the remaining 30\% as the validation set.  
The predictive abilities of each joint model for the 3-month risk of death were assessed using repeated measures of CD4 levels collected at landmark times of \( s = 0, 3, 6, 9,\) and \(15\) months. 
\\
This assessment utilized the AUC, BS, as well as their integrated counterparts (iAUC and iBS), to compare the dynamic predictions of the ZINB and ZIP joint models applied to the HIV data (see Figure \ref{auc1} and Table \ref{auc2}). \\
Table \ref{auc2} summarizes the time-dependent predictive performance of the ZINBJMCF and ZIPJMCF models in the HIV cohort, using AUC and Brier Score (BS) at landmark times. The table also reports the integrated metrics (iAUC and iBS) that summarize predictive ability over the full follow-up period.
\\
The AUC section shows the discrimination ability of each model at each landmark time, with higher values indicating better differentiation between patients who experience the event and those who do not. The ZINBJMCF model achieves AUCs ranging from 0.846 to 1.000, whereas the ZIPJMCF model ranges from 0.867 to 1.000. The corresponding p-values indicate no statistically significant differences at most landmark times. The integrated AUC (iAUC) over the study period is slightly higher for ZIPJMCF (0.909) compared to ZINBJMCF (0.884), summarizing overall discrimination across all times.
\\
The Brier Score section evaluates predictive accuracy and calibration, with lower values indicating better agreement between predicted and observed outcomes. Both models show modest increases in BS over time, with values ranging from 0.061–0.100 (ZINBJMCF) and 0.062–0.101 (ZIPJMCF). Integrated Brier Scores (iBS) are similar between the models (0.086 vs. 0.087), reflecting comparable overall calibration.
\\
The corresponding Figure \ref{auc1} visually illustrates these AUC and BS values over landmark times, highlighting the evolution of model performance. It provides an intuitive depiction of how discrimination and accuracy change across the follow-up, complementing the numerical summary in the table.\\

\begin{figure}[hbt!]
\centering
\includegraphics[width=0.45\textwidth]{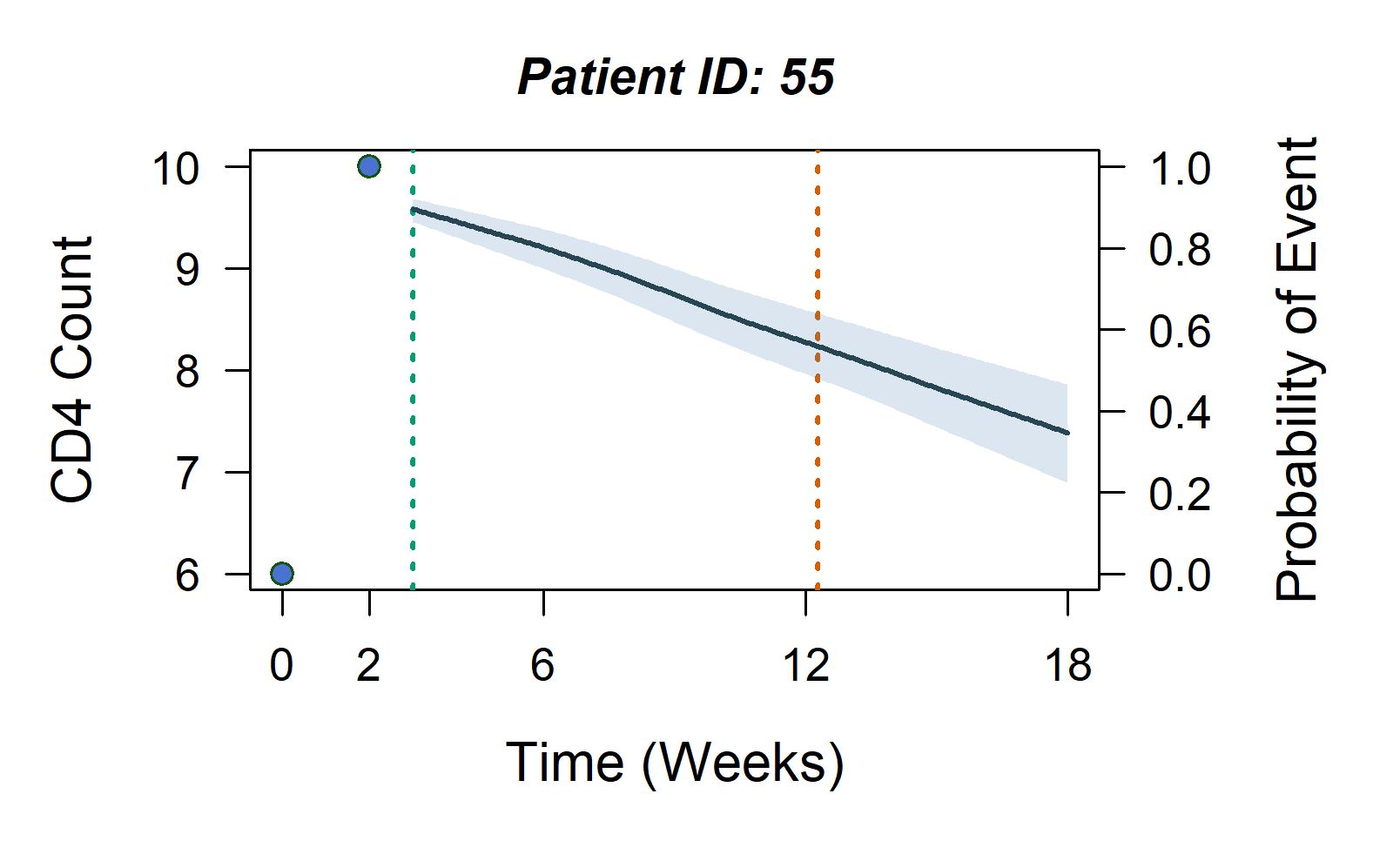}
\includegraphics[width=0.45\textwidth]{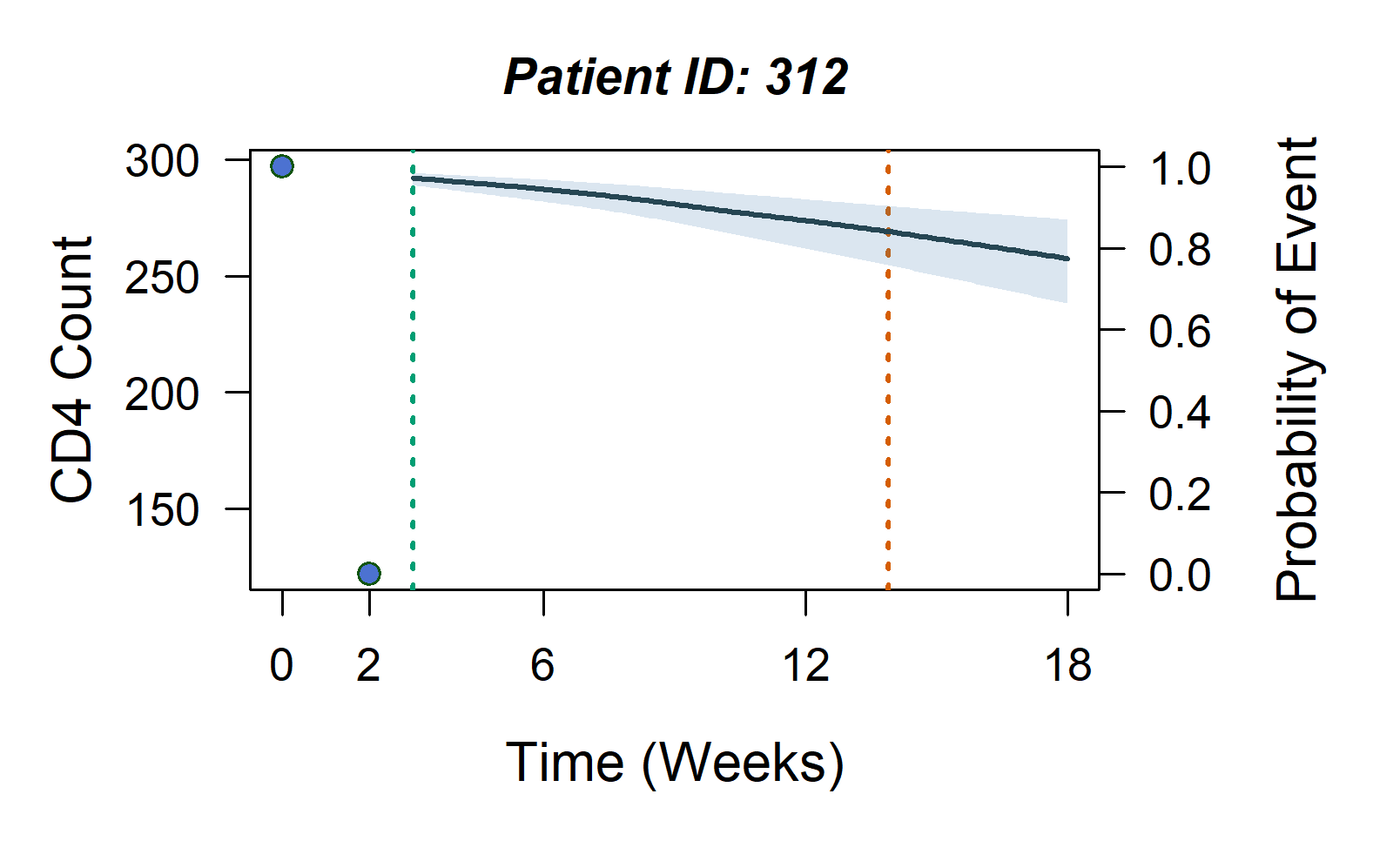}\\[0.3cm]
\includegraphics[width=0.45\textwidth]{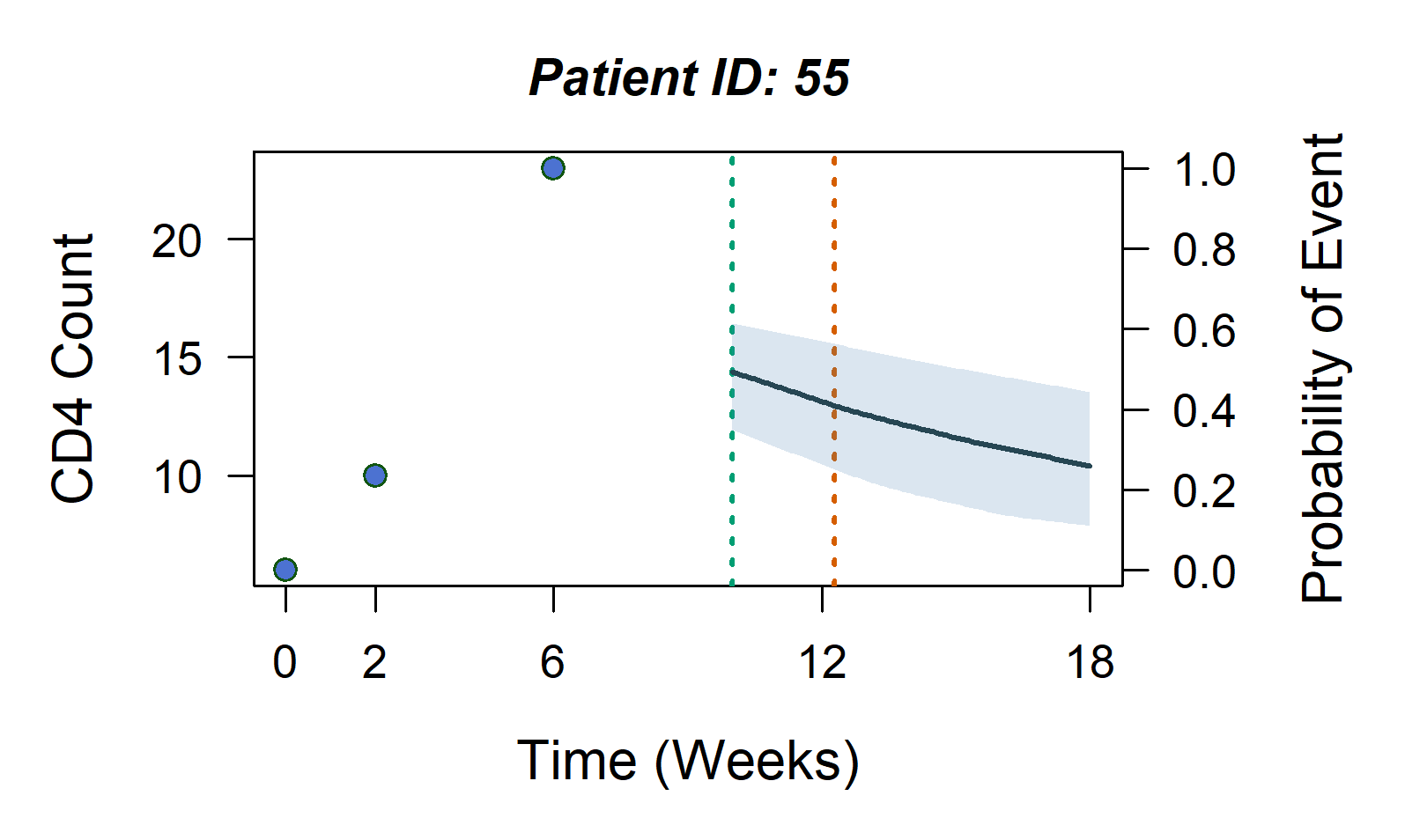} 
\includegraphics[width=0.45\textwidth]{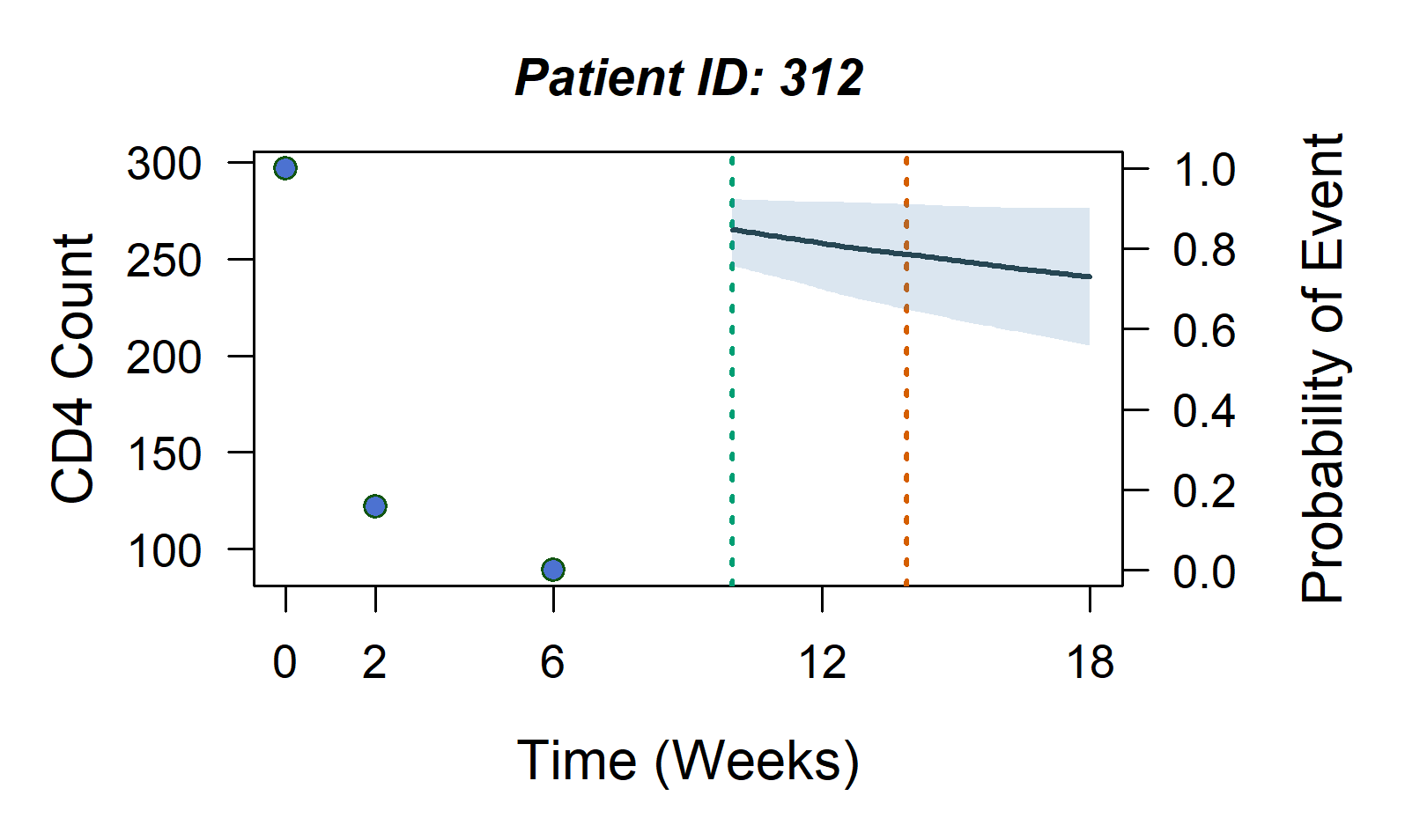}
\vspace*{-.1cm} 
\caption{\label{pred_ind} 
Death prediction, expressed as $1-\pi_i(s+\text{Time} \mid s)$, at the landmark times $s=3$ and $s=10$ weeks (indicated by the green dashed lines), along with the corresponding 95\% credible intervals, for two randomly selected patients who are at risk at these landmark times based on ZINBJMCF model. The red dashed lines indicate the observed survival times.
}

\end{figure}

Figure \ref{pred_ind} presents the dynamic death prediction 
$1 - \pi_i(s + \text{Time} \mid s)$
at the landmark times \( s = 3 \) and \( 10 \) weeks, along with the corresponding 95\% credible intervals, for two randomly selected patients (IDs 55 and 312) who are at risk at these time points. These predictions are based on the best-fitting ZINB joint model incorporating a cure fraction. The figure illustrates the reliability of these dynamic predictions. Specifically, patient 55, with very low CD4 counts, has a higher predicted risk of death compared to patient 312, whose higher CD4 counts correspond to a substantially greater chance of survival.\\
Furthermore, the estimated cure probabilities based on PrevOI highlight significant differences: patients without prior opportunistic infections (PrevOI=0) have an estimated cure probability of approximately 0.198, whereas patients with prior opportunistic infections (PrevOI=1) exhibit a substantially lower cure probability of about 0.006. This finding suggests that previous opportunistic infections are strongly associated with a reduced chance of cure, which aligns with the survival predictions illustrated in Figure \ref{pred_ind}.
It is important to note that this interpretation does not account for the standard deviation of the model parameters, which results in wide credible intervals. This uncertainty is primarily due to the limited sample size, the inherent variability in the patient population, and a large rate of missingness in the data, all of which contribute to less precise estimates of the cure fraction.

\begin{table}[ht]
\centering
\caption{\label{auc2} Comparison of time-dependent AUC and Brier Score (BS), reported as mean $\pm$ SD, between the ZINBJMCF and ZIPJMCF models.}
\begin{tabular}{c|ccc}
\hline
\multicolumn{4}{c}{AUC} \\
\hline
Time & ZINBJMCF & ZIPJMCF & p-value \\
\hline
0  & 0.871 $\pm$ 0.061 & 0.891 $\pm$ 0.063 & 0.818 \\
3  & 0.924 $\pm$ 0.033 & 0.959 $\pm$ 0.017 & 0.346 \\
6  & 0.846 $\pm$ 0.062 & 0.867 $\pm$ 0.059 & 0.812 \\
9  & 0.866 $\pm$ 0.053 & 0.897 $\pm$ 0.056 & 0.689 \\
15 & 1.000 $\pm$ 0.000 & 1.000 $\pm$ 0.000 & 1.000 \\
\hline
iAUC & 0.884 $\pm$ 0.004 & 0.909 $\pm$ 0.003 &   \\
\hline
\multicolumn{4}{c}{Brier Score (BS)} \\
\hline
Time & ZINBJMCF & ZIPJMCF & p-value \\
\hline
0  & 0.061 $\pm$ 0.014 & 0.062 $\pm$ 0.014 & 0.964 \\
3  & 0.067 $\pm$ 0.015 & 0.068 $\pm$ 0.015 & 0.978 \\
6  & 0.085 $\pm$ 0.019 & 0.085 $\pm$ 0.019 & 0.994 \\
9  & 0.100 $\pm$ 0.018 & 0.101 $\pm$ 0.018 & 0.979 \\
15 & 0.086 $\pm$ 0.020 & 0.087 $\pm$ 0.020 & 0.972 \\
\hline
iBS & 0.086 $\pm$ 0.002 & 0.087 $\pm$ 0.002 &   \\
\hline
\end{tabular}
\end{table}

\FloatBarrier

\section{Conclusion and Discussion}
This study proposed a comprehensive Bayesian joint modeling framework for zero-inflated longitudinal count data and time-to-event outcomes in the presence of a cure fraction. The approach integrates a flexible mixed-effects hurdle model for the longitudinal process with a Cox proportional hazards mixture cure model for the survival process, enabling simultaneous modeling of excess zeros, overdispersion, and long-term survival probabilities. By employing Hamiltonian Monte Carlo and Variational Bayes, we achieve efficient inference despite the high dimensionality and complexity of the model, while maintaining flexibility in linking longitudinal and survival components.
\\
The application to an HIV cohort demonstrated clear practical advantages. In particular, modeling overdispersion via a zero-inflated negative binomial (ZINB) distribution provided better model fit and more accurate dynamic predictions than a zero-inflated Poisson (ZIP) alternative. The ZINB-based joint model offered superior short-term mortality risk discrimination and calibration, especially at later landmark times, and produced clinically interpretable cure probabilities aligned with known prognostic factors, such as the presence of prior opportunistic infections.
\\
Simulation studies further confirmed the robustness of the proposed framework, with near-unbiased parameter estimates, nominal coverage rates, and reliable dynamic risk predictions, particularly in overdispersed settings. In contrast, models that ignored overdispersion displayed substantial bias and poorer predictive performance, highlighting the importance of appropriate longitudinal distributional assumptions.
\\
In this work, we adopted the {current value} parameterization to link the longitudinal and survival processes. However, other association structures, such as slope-based effects, cumulative biomarker history, or shared latent processes \citep{ganjali2024joint}, could be valuable extensions, offering richer insight into the temporal relationship between biomarker dynamics and event risk. Incorporating these alternatives represents a promising direction for future research.
\\
Overall, the proposed zero-inflated joint modeling framework with a cure fraction broadens the methodological toolkit for analyzing complex biomedical data. Its ability to generate individualized, dynamically updated risk predictions makes it well-suited for precision medicine and clinical decision support. Potential future developments include extending the model to cover all responses belong to exponential dispersion family \citep{ganjali20252}, 
to multivariate longitudinal outcomes, accommodating competing risks, integrating high-dimensional biomarkers, and implementing the methodology in real-time prognostic tools for clinical use.

\bibliography{sample}

\end{document}